# Unlocking High Performance, Ultra-Low Power Van der Waals Transistors: Towards Back-End-of-Line In-Sensor Machine Vision Applications

*Olaiyan Alolaiyan[1,†], Shahad Albwardi[1,†], Sarah Alsaggaf[1,†], Thamer Tabbakh[1,4], Frank W. DelRio[2], and Moh. R. Amer*[1,3,4]*

[1]Center of Excellence for Green Nanotechnologies,
Microelectronics and Semiconductor Institute
King Abdulaziz City for Science and Technology, Riyadh, 11442, Saudi Arabia
[2]Sandia National Laboratories,
Material, Physical, and Chemical Sciences Center, Albuquerque, NM, 87123, USA
[3]Department of Electrical and Computer Engineering
University of Southern California, Los Angeles, CA, 90089, USA
[4]Department of Electrical and Computer Engineering
University of California, Los Angeles, Los Angeles, CA, 90095, USA

[†]These authors contributed equally
*Please send all correspondence to mamer@seas.ucla.edu, mamer@kacst.edu.sa,

**Abstract:**
Recent reports on machine learning (ML) and machine vision (MV) devices have demonstrated the potentials of 2D materials and devices. Yet, scalable 2D devices are being challenged by contact resistance and Fermi Level Pinning (FLP), power consumption, and low-cost CMOS compatible lithography processes. To enable CMOS+2D, it is essential to find a proper lithography strategy that can fulfill these requirements. Here, we explore modified van der Waals (vdW) deposition lithography and demonstrate a relatively new class of van-der-Waals-Field-Effect-Transistors (vdW-FETs) based on 2D materials. This lithography strategy enables us to unlock high performance devices evident by high current on-off ratio ($I_{on}/I_{off}$), high turn-on current density ($I_{on}$), and weak Fermi Level Pinning (FLP). We utilize this approach to demonstrate a gate-tunable near-ideal diode using $MoS_2/WSe_2$ heterojunction with an ideality factor of ~1.65 and current rectification of $10^2$. We finally demonstrate a highly sensitive, scalable, and ultra-low power phototransistor using $MoS_2$/ $WSe_2$ vdW-FET for Back-End-of-Line (BEOL) integration. Our phototransistor exhibits the highest gate-tunable photoresponsivity achieved to date for white light detection with ultra-low power dissipation, enabling ultra-sensitive, ultra-fast, and efficient optoelectronic applications such as in-sensor neuromorphic machine vision. Our approach shows the great potential of modified vdW deposition lithography for back-end-of-line CMOS+2D applications.

## INTRODUCTION:

Sensory nodes are becoming in significant demand with the rise of Artificial Intelligence (AI) and big-data processing. In fact, it is predicted that a rapid growth in integrated sensors technologies will take place in the next few years. Driven by the intuitive desire to implement machine learning (ML) applications, on chip AI/ML is becoming a vital part for many industries. However, current optical sensor technologies are limited by conventional van Neumann architecture, where the sensing element and the computing element are physically separated. A new approach for the future of semiconductors is readily proposed consisting of CMOS+X where X is an emerging technology [1]. Within this context, multifunctional sensors with in-sensor and near-sensor computing have emerged as the forefront for efficient AI/ML applications [2-5].

Accordingly, in-sensor Intelligent Machine Vision (IMV) has the potential to revolutionize many industries such as autonomous vehicles, robotics, and surveillance. Inspired by the human eye, IMV requires pre-processing of collected information within the sensory node and without the need for ADC. Although near sensor and in-sensor machine vision are both attractive approaches, in-sensor machine vision exhibit more benefits than near-sensor machine vision including lower number of devices over a limited size, less fabrication processes, and perhaps lower power consumption. Nevertheless, most of reported work on in-sensor machine vision suffer from limitations in the photodetector element which include low performance, scalability issues, expensive or impractical fabrication processes, and noticeable power consumption. These major challenges prompt exploring a reliable photodetector element first that can be utilized for in-sensor machine vision, exhibit high performance in the dedicated detection wavelength, ultra-low power consumption, and back-end-of-line compatible.

Luckily, 2-Dimensional materials are becoming a strong candidate for such applications [6-9]. Various reports have shown the capability of using 2D materials for in-sensor and near sensor machine vision. Due to their atomically thin structure coupled with exceptional electronic and optical properties, 2D materials are a promising candidate for intelligent machine vision. Yet, a major bottleneck that prevents 2D devices from being implemented in such an application is device resistance which suffers from FLP at the contacts. This challenge can affect photodetector and phototransistor properties including the photoresponsivity magnitude and its electrostatic (gate) tunability, which are some major features required to implement in-sensor machine vision concept.

In this work, we demonstrate a new and scalable strategy to fabricate 2D-FET phototransistors based on modified vdW deposition lithography. We introduce vdW-FET devices and show device resistance can be mitigated by a newly developed treatment process called “anneal and heal,” which produces relatively high performance vdW-FETs. We then explore $MoS_2$/ $WSe_2$ junction in vdW-FET configuration and show gate-tunable, near-ideal diode behavior. We finally explore optoelectronic properties of this phototransistor and demonstrate highly tunable photoresponsivity with ultra-low power dissipation, superseding other reported photodetector devices used for in-sensor machine vision applications.

## RESULTS:

### Conventional 2D-FETs vs. VdW-FETs:

One of the major hurdles that prevents 2D-FETs from CMOS integration is the device resistance. Conventional metal evaporation on 2D materials can induce structural damage and defects on the targeted 2D material, which causes Metal Induced Gap States (MIGS) leading to FLP. Figure 1a demonstrate the energy band diagram of a conventional $MoS_2$ device fabricated using this method [10]. Nevertheless, recent investigations by different groups have shown that it is possible to significantly reduce MIGS and FLP by using semi-metals such as Bismuth (Bi) or Antimony (Sb) with $(01\overline{1}2)$ orientation [11, 12]. Although the use of these semi-metals is a promising approach for ultralow contact resistance, there is ambiguity lingering in terms of their manufacturing scalability, cost effectiveness, and their integration for in-sensor computing/machine vision.

In contrast, vdW-FET structure is shown in figure 1b. Here, electrodes are patterned using optical lithography first with 2D material deposited on top of source and drain electrodes, as shown schematically in the figure. Due to the dangling-free bonds on the surface of 2D materials, the interface between 2D material and the metal electrode is van der Waals (hence the name). Here, the energy band diagram after 2D deposition shows a large Schottky barrier, leading to a non-linear current-voltage characteristics and hence, a large device resistance. This behavior is shown in the electrical characteristics of $MoS_2$ vdW-FETs in figures 2b and 2c (red curves) for $IV_{ds}$ and $IV_{gs}$, respectively. In fact, some devices show no current flow (open circuit) in the channel as shown in figure S1a. This high device resistance is attributed to the weak bonding between metal electrodes and $MoS_2$. It also stems from the nature of vdW alignment of Au metal work function to the middle of the gap of $MoS_2$, as shown in figure S2 and discussed elsewhere [13].

To mitigate this high resistance effect, we developed a new scalable treatment strategy using pulsed thermal annealing technique. Here, fabricated vdW-FET devices are exposed to a thermal pulse with temperatures of 350$^{o}$C for few seconds (usually 2-5 seconds) followed by rapid cooling, as demonstrated schematically in figure 2a (See methods for more information). The results of anneal and heal method are shown in figures 2b and 2c (blue curves) for $IV_{gs}$ and $IV_{ds}$ of $MoS_2$ vdW-FET device, respectively. A dramatic increase in the device electrical performance is observed. The measured low bias resistance becomes linear. In fact, for large channel devices, applying higher $V_{ds}$ will resolve current saturation as shown in figure S5. The observed $I_{on}$ in figure 2b is enhanced by an order of at least $10^3$. We attribute this current enhancement to the strong coupling between the contact resistance and metal electrodes after anneal and heal treatment process.

Our developed treatment process is also effective on other 2D materials such as $WSe_2$ and black phosphorus. This is demonstrated in figures 2e-2f. Here, $IV_{gs}$ before and after anneal and heal treatment process show enhanced device performance for $WSe_2$ and black phosphorus. The maximum current density $I_{on}$ and current on-off ratio $I_{on}/I_{off}$ show dramatic increase after anneal and heal treatment process. That is, $I_{on}$ increases by 3270 for $WSe_2$ and 52 for BP, while $I_{on}/I_{off}$ ratio observed increases from 0 to $10^4$ for $WSe_2$ and 8.1 to $10^2$ for black phosphorus. Additional measurements and devices are illustrated in figures S3 and S4 for $WSe_2$ and black phosphorus, respectively.

**Performance of vdW-FETs:**

To elucidate on vdW-FET device performance after anneal and heal treatment process, we carried out systematic measurements of different channels using $MoS_2$ vdW-FETs, as shown in figure 3a-3c. 3 different devices were fabricated with different channel lengths, mainly 2um, 6um, and 9um. For all devices, the gate electrode and dielectric thickness are kept constant (300nm). We observe a monotonic decease in $I_{on}$ with increasing channel length. This trend is expected since increasing channel length will increase free carriers scattering effects and hence, lower attainable current. Additional electrical characterization showing current saturation is demonstrated in figure S5 along with device resistance vs. channel length of different vdW-FET devices.

It is worth mentioning that almost all $MoS_2$ devices reported in literature show *n*-type conductivity regardless of the metal type used [10, 14]. This behavior is attributed to strong Fermi Level Pinning (FLP) caused by Metal Induced Gap States (MIGS). Ideally in a van der Waals metal-semiconductor junction, different metals with different work function (φ) should give rise to different conductivity profiles, which is attributed to alignment of the metal work function to the valance or the conduction band [13]. For instance, for Pt electrodes ($\varphi \approx 5.6$ eV) with $MoS_2$ nanosheets, the valance band aligns with Pt work function, giving rise to *p*-type conductivity. This behavior has been previously observed in vdW contacts [10].

To shed some light on the nature of our vdW-FET contacts after anneal and heal treatment process, we fabricated $MoS_2$ devices with high work function metal (Pt). In figure 3d, we show $IV_{gs}$ of $MoS_2$ device with Pt electrodes before and after anneal and heal treatment process. Before any treatment, the device exhibits low current, analogous to Au- $MoS_2$ devices. After anneal and heal treatment, however, the device exhibits strong *p*-type profile with improved $I_{on}/I_{off}$ ratio reaching$10^2$ and enhanced turn-on current by at least 32 folds. Figure S6 shows additional vdW-FET devices with Pt electrodes displaying similar trend. This observed *p*-type conductivity of vdW-FET is in contrast with Pt- $MoS_2$ devices fabricated via metal evaporation method which predominantly shows *n*-type behavior due to strong FLP. We attribute this observed *p*-type behavior to evidence of weak FLP and clean interface between the metal and 2D material after anneal and heal treatment process.

Our vdW-FETs based on $MoS_2$ exhibit relatively competitive device performance compared to those in literature. This is demonstrated in figure 3e where highest $I_{on}$ is plotted at different channel lengths and compared to other high performance $MoS_2$ devices [11, 12, 15-18]. The shaded pink area shows the expected $I_{on}$ region of vdW-FETs with different channel lengths. This area also overlaps with other reported high-performance $MoS_2$ 2D-FET devices with different metals. What distinguishes vdW-FET devices here is the simplicity of fabricating these vdW-FETs which yield cost-effective solutions and scalable platform for 2D-FETs. In addition, vdW-FETs exhibit weak FLP with conventional metals, a drawback found in many reported $MoS_2$ transistors. Thus, we believe this modified vdW deposition lithography strategy is appealing to produce high performance scalable phototransistors and photodetectors for different back-end-of-line optoelectronic applications.

**Electron Transport of $MoS_2/WSe_2$ vdW-FET with near unity ideality factor:**

To gain a deeper understanding of modified vdW deposition lithography capabilities, we demonstrate a *pn* heterojunction using $MoS_2/WSe_2$ in vdW-FET configuration, as shown in the schematics in figure 4a. Here, $MoS_2$ and $WSe_2$ nanosheets are deposited first followed by the anneal and heal treatment process for multiple cycles. Figure 4b shows the measured $IV_{ds}$ before and after anneal and heal treatment process (see figure S7 for more details). the measured $IV_{ds}$ curve readily after 2D deposition shows very poor rectification profile with current varying between 1pA to 0.1nA. However, after applying anneal and heal treatment process, the device performance improves dramatically showing a diode behavior with a near-unity ideality factor ($n = 1.65 \pm 0.1$) and current rectification ratio of $10^2$. figure 4c shows the fitted curve to the ideal diode equation using $I = I_s \left[exp\left(\frac{eV_{ds}}{nkT}\right) - 1\right]$ where $I_s$ is the reverse saturation current, $T$ is the temperature, $e$ is the electric charge, and $k$ is Boltzmann constant. Figures 4d and 4e plot the evolution of current rectification ratio and ideality factor with anneal and heal treatment time, respectively.

Apart from its fabrication simplicity with near unity ideality factor and high current rectification, what makes these $MoS_2/WSe_2$ vdW-FET diodes interesting is their gate-tunable capability. Figure 4f demonstrate this gate tunable response before and after anneal and heal treatment process. Before any treatment, device exhibits extremely high resistance with very low current. However, after anneal and heal treatment process, we see two distinctive resistive regions, high resistance region and low resistance region. The high resistance region occurs when the device is gated in the negative regime (-10V > Vgs > -4V), giving rise to current density lower than 0.01 ($A/cm^2$), while the low resistive region occurs when the gate voltage varies between -4V< Vgs < 10V, with a maximum current density up to 0.78 ($A/cm^2$). The resistance change between these regions is in the order of 44 folds, offering a large switching range. This gate tunable behavior observed in $MoS_2/WSe_2$ vdW-FET can be desirable for many low power switching applications.

**Highly sensitive gate-tunable Optical Photodetector using $MoS_2/WSe_2$ vdW-FET:**

To enable in-sensor machine vision concept and perform multiply and accumulation operations (MAC), an array of reconfigurable sensors with $m \times n$ elements can be connected as shown in figure S8. Each sensor element has responsivity ($R_{mn}$). with light being the external stimulation with power ($P$), it is possible to express the summation current as:

$$I_n = \sum_{n=1}^{N} I_{mn} = \sum_{n=1}^{N} R_{mn} P_n \quad (1)$$

Here, the multiplication in equation 1 is performed at the individual sensor level. Accordingly, it is essential that each individual sensor needs to exhibit tunable photoresponsivity (or responsivity hereafter) with external modulation such as gate voltage, which emulates the synaptic weight of a neural network. To our knowledge, very few reports have demonstrated in-sensor machine vision using this principle. Most notable is the work reported by Mennel *et al.* where they demonstrated in-sensor machine vision using self-powered split-gate WSe2 photodetector in a floating gate configuration [6]. The split gate WSe2 photodetector exhibit tunable responsivity by changing the electrostatic doping. Nevertheless, most reports utilize near-sensor computing scheme instead of in-sensor computing as will be discussed later.

One can deduce that optimizing the photodetector requirements is a vital step towards CMOS integrable in-sensor machine vision. As such, in addition to the tunable responsivity requirement and to enable multifunctional optical sensors on CMOS platform for machine vision applications, any photodetector element should exhibit the following requirements: (1) scalable and cost-effective fabrication techniques for back-end-of-line integration on CMOS platform (2) reconfigurable high responsivity with high degree of tunablility using external voltage (high gate tuning ratio). (3) white light detection to enable a wide range of wavelength detection, and most importantly, (4) ultra-low power consumption. Accordingly, identifying a photodetector that can meet all these requirements is a vital step before integrating any photodetector element for in-sensor machine vision (i.e. using floating gate structure).

To date, no photodetector devices used for in/near-sensor computing can meet all these stringent requirements. In fact, most published reports use methods and techniques that either hard to scale up or CMOS incompatible [19, 20]. These methods include the use of electron beam lithography, non-conventional substrates, or expensive fabrication processes [21-23]. Moreover, all reports show devices with noticeable power dissipation which can affect the total performance [7, 20, 24-28]. Other reports show photodetectors with low responsivity or low gate tunability [29-34]. Additionally, most reports on photodetectors utilized for in/near-sensor computing are aimed for specific wavelength detection and not the entire visible range.

Using modified vdW deposition lithography, we demonstrate one-of-a-kind photodetector that can fulfill the requirements stated above. Here, the photodetector utilizes $MoS_2/WSe_2$ junction fabricated on $SiO_2/Si$ substrate, as shown in figure 5a. The measured $IV_{ds}$ characteristics before and after LED white light illumination ($P$ = 8mW/cm$^2$) are shown in figure 5b and 5c for $V_{gs}$=5V and $V_{gs}$= -19V, respectively. The inset of each figure shows the obtained photocurrent ($I_{ph} = I_{light} - I_{dark}$) plotted against $V_{ds}$. Under $V_{gs}$= 5V, the highest photocurrent magnitude is obtained with the maximum occurring in the reverse bias. However, when the applied gate voltage is Vgs = -19V, the photocurrent is remarkably suppressed by 9.68 and 6.625 orders of magnitude in the reverse bias (at $V_{ds}$= -100mV) and at zero $V_{ds}$, respectively. This photocurrent modulation is a key feature in obtaining gate tunable responsivity for in-sensor machine vision.

To gain a deeper understanding of the $MoS_2/WSe_2$ photodetector performance, we show the obtained photoresponsivity (sometimes referred to as responsivity) and specific detectivity of the device at different applied $V_{ds}$ and $V_{gs}$ in figures 5d and 5e, respectively. Figures S9b and S9c show some sampled $IV_{ds}$ curves in dark and under white light illumination, respectively. The highest responsivity occurs when the device is reverse biased (at $V_{ds}$ =100mV) and gated at $V_{gs}$ = 5V. In contrast, the highest detectivity is obtained when the device is self-powered ($V_{ds}$ = 0V) and when gated at $V_{gs}$ = 5V. This can be clearly seen in figures S9d and S9e where the responsivity and detectivity are plotted for $V_{ds}$ =0V and $V_{ds}$ =-100mV, respectively.

Based on these results, one can deduce that the photodetector can operate under 2 different modes, photovoltaic mode and photoconductive mode. In photovoltaic mode (or self-power photodetector), the applied $V_{ds}$ is zero and a short circuit current ($I_{sc}$) is generated under white light illumination. The rise and fall time of the photodetector under this mode is shown in figure S10 and is estimated to be 25ms and 8ms, respectively. However, when reverse voltage is applied, additional photogenerated carriers are created leading to higher photocurrent. In both cases, the responsivity can be tuned by 6.6 folds for photovoltaic mode and 9.7 folds for photoconductive mode, as shown in figures S9c and S9d. Here, the device exhibits high gate tuning ratio with 38.2mA/V and 53.5mA/V for photovoltaic mode and photoconductive mode, respectively.

The power dissipation of this device is shown in figure S11 at different $V_{ds}$ and $V_{gs}$. Here, power dissipation is defined as $P_d=I_{dark}V_{ds}$ at any given $V_{gs}$. Under photovoltaic effect, the dissipated power is zero and a short circuit current is generated (power generation instead of power dissipation). However, under photoconductive mode, the power dissipation is not zero and high dark current can lead to high power dissipation. Thus, to optimize power dissipation, devices operating under photovoltaic mode with tunable responsivity are favorable for in-sensor machine vision.

The performance of our device supersedes reported photovoltaic devices when operating under photovoltaic mode, as shown in the benchmark plot in figure 6a. Here, the device performance (Responsivity vs. $V_{gs}$) is plotted from reports utilizing gate-tunable photovoltaic devices for machine vision [6, 21, 23, 29, 31-33, 35]. The highest responsivity reported to date ranges between 0.05 and 0.06 (A/W). These reports utilize a specific wavelength detection while white light machine vision photovoltaic devices show responsivities that ranges between 0.009 and 0.0137 (A/W). In contrast, our device shows superior responsivity with the highest reaching 0.45 (A/W) with high gate tunability.

The power dissipation parameter is a vital factor for scalable in-sensor or near sensor machine vision applications. In figure 6b, we benchmark our device with prominent machine vision photodetectors reported in literature by plotting the responsivity vs. dissipated power [7, 19-22, 24-28, 34, 36-38]. Here, various photodetector technologies reported for machine vision are being compared to our device including InP, $CsPbBr_3$ perovskite, $BiFeO_3$, etc. White light photodetectors are also included in this benchmark and plotted as unfilled dots with crossed marks.

Based on this, our device surpasses all photovoltaic devices analogues to figure 6a. More importantly, when our device operates under photoconductive mode, the obtained device performance shows high responsivity with very low power dissipation. The highlighted area in pink of figure 6b shows the region of operation of our phototransistor operating under photodetector mode. This region is comparable to commercial silicon photodiodes with very low dark current. While other reports show high responsivities, they all fail in the power dissipation requirement. This is due to the required bias voltage needed to obtained high photogenerated carriers which is associated with a noticeably large dark current. Thanks to modified vdW deposition lithography, efficient photogenerated carrier extraction is achieved using $MoS_2$/$WSe_2$ in vdW-FET configuration.

Although these reports show noticeably higher power dissipation compared to our device, they also fail to create a scalable and low-cost fabrication processes. As stated above, most reports use electron beam (e-beam) lithography or other means of complicated or expensive lithography processes, which can hinder the integration of these devices for large scale applications. Yet, authors understand that several groups use this method to demonstrate proof of concept devices only.

The use of optical lithography coupled with modified vdW deposition lithography is a desirable route to produce scalable, high-performance, and cost-effective 2D transistors. In principle, we believe this vdW deposition lithography technique can be applied to wafer scale grown 2D materials such as CVD grown $MoS_2$ and $WSe_2$. Here, wafer scale 2D semiconductors need to be transferred to our patterned substrate. Various defect-free transfer methods have been investigated in literature [39]. Such work will be investigated in the future to perform high-level image processing. Nonetheless, the work presented here serves as the backbone for a proof-of-concept photodetector based on modified vdW deposition lithography for BEOL machine vision, and will be used for comparison with CVD grown 2D materials.

## CONCLUSIONS:

We have fabricated vdW-FETs using modified vdW deposition lithography. This strategy enables us to create low-cost, scalable, energy efficient, and high performance BEOL phototransistors for in-sensor machine vision applications. We showed that vdW deposition lithography alone cannot produce high performance devices due large device resistance attributed to weak bonding between 2D material and metal electrodes. However, incorporating anneal and heal treatment process will produce low device resistance with high current density. This lithography technique enables us to eliminate FLP, which is a major challenge in 2D-FETs. This strategy is utilized to produce gate-tunable near-ideal diode behavior using $MoS_2$/$WSe_2$ vdW-FET with ideality factor of 1.65 and current rectification of $10^2$. Finally, we demonstrate one-of-a-kind gate-tunable photodetector using $MoS_2/WSe_2$ vdW-FET. This photodetector can operate in 2 different modes, photovoltaic mode and photoconductive mode. Under photovoltaic mode, the device exhibits gate-tunable responsivities reaching 0.45 (A/W) with responsivity modulation of 6.6 folds. When operating under photoconductive mode (reverse bias), the photodetector shows responsivities up to 0.68 A/W with responsivity modulation of 9.68 folds. This photodetector performance is benchmarked with other reported photodetectors used for in-sensor and near-sensor machine vision applications. We deduce that our photodetector exhibits superior performance compared to other reported photodetectors utilized in machine vision applications by considering the requirements for in-sensor machine vision. Broadly, our work demonstrate a novel method to create BEOL 2D transistors for CMOS compatible multifunctional sensor applications, enabling CMOS+2D.

## METHODS:

### Device Fabrication:

VdW-FETs devices are fabricated by using $SiO_2$ (300nm) grown on silicon substrate. Source and drain electrodes are first patterned using optical lithography followed by 100nm electrode deposition. 2D nanosheets (from 2D semiconductor) are mechanically exfoliated using special exfoliation tape followed by micro-alignment to the pre-patterned substrate using a customized dry transfer setup. Optical inspection, Raman measurements, and *IV* measurements were carried out to confirm the contact of the deposited nanosheet with the source and drain electrodes. The substrate was thoroughly cleaned with acetone, IPA, and deionized water followed by nitrogen flow to ensure the cleanliness of the surface prior to 2D material deposition. Figure S12 shows the fabrication process schematically.

### Anneal and Heal Treatment Process:

Fabricated chips are placed on top of a hot plate in ambient for 2-5 seconds where a global thermal pulse is applied. Chips are removed directly after the annealing cycle and left at room temperature for few seconds to cool down. Optical examination of the substrate before and after annealing does not show any changes to the substrate nor the deposited 2D material. However, repeating the annealing cycles for a prolonged period of time will produce degraded device performance and damage to the deposited 2D material, evident by observation of defects on the 2D surface via optical microscope.

For $MoS_2$-Au devices here, gold electrode work function falls in the mid-gap of $MoS_2$ with no band alignment. This causes no current flow as observed in many devices before any treatment. The anneal and heal treatment process seems to eliminate barriers between the metal and 2D material including defects. However, the origin of the full mechanism of alignment with the Conduction Band Minimum (CBM) is not fully understood yet. One possible explanation is that due to substrate doping effect caused by trapped charges in $SiO_2$, it is favorable for $MoS_2$ CBM to align with Au work function, giving rise to a strong vdW contact with *n*-type conductivity. In fact, when fabricating devices on a different dielectric ($Al_2O_3$), it is possible to get *p*-type conductivity as shown in figure S13, which is further evidence that of induced doping of $SiO_2$. It also demonstrate that anneal and heal treatment process produces weak FLP. Yet, further analysis into the dynamics of band alignment after anneal and heal treatment process should be explored in the future.

Unlike $MoS_2$ devices, where the maximum device performance occurs readily after the first treatment cycle, other 2D materials such as $WSe_2$ and black phosphorus show device enhancement occurring after multiple treatment cycles, as shown in figures S3-S4.

**Optical and Morphology Characterization:**
Each sample is characterized using confocal Raman spectroscopy (Renishaw) with a silicon detector and 2400 1/mm grating. We used 532-nm laser line to excite Raman modes of our 2D materials. To avoid overheating, we use laser power up to 0.93mW for a laser spot size of ~1.5µm. 100× objective lens was used for all Raman measurements. Morphology measurements were carried out using Park systems XE7. We use non-contact mode to ensure no induced defects/changes to the deposited nanosheets on our devices. Optical and AFM images were compared before and after measurements to make sure the device is still intact without any structural changes.

**Electrical characterization:**
Current-Voltage characteristics were measured using Keysight B1500 with each of source, drain, and gate electrodes connected to the 3 different micromanipulators. $IV_{ds}$ and $IV_{gs}$ measurements were carried out using current sensitive modules (down to 1fA sensitivity). Measurements were carried out in room temperature and in ambient conditions. Before and after each electrical measurements, optical inspection and Raman measurements were carried out to identify any change in the structure.

**Photodetector Characterization:**
Photo-*IVs* under dark and light illumination was carried out in ambient using our probe station setup. The chip was probed prior to any annealing and *IV* measurements under dark and LED white light (OMAX20W) illumination was measured. $V_{ds}$ and $V_{gs}$ are varied to obtain current maps. Anneal and heal treatment process is subsequently applied to the chip and the photo-*IV* measurements under dark and white light illumination are readily carried out. This process is repeated for every anneal and heal treatment cycle until a lower value in the maximum current is obtained.

**Responsivity and Detectivity Calculations:**
The responsivity ($R$) is given by ( $R = I_{ph}/P_{op}$), where $I_{ph}$ is the photocurrent (in units of mA/cm$^2$) and $P_{op}$ is the optical power incident on the photodetector (in units of mW/cm$^2$). The calculated specific detectivity ($D^*$ in units of Jones) of the photodetector is given by $D^* = RA^{\frac{1}{2}}/(2eI_{dark})^{1/2}$, where $A$ is the effective area of the junction. To get accurate responsivity values, the optical power density of the LED white light is characterized using an optical power meter (Thorlabs PM400 and S120C). The power density vs. wavelength is characterized as shown in figure S14. The total power density of the white light is measured using irradiance mode (in units of mW/cm$^2$) at the exact location of the device. The obtained power density is found to be 8mW/ cm$^2$.

**ACKNOWLEDGEMENTS:**

This research was financially supported by King Abdulaziz City for Science and Technology (KACST) through the Center of Excellence for Green Nanotechnologies (CEGN), award 20132944. T.A.T and M.R.A would like to acknowledge the support of KACST cleanroom core lab facilities. F.W.D. would like to acknowledge the support of the Center for Integrated Nanotechnologies, a Department of Energy office of Basic Energy Sciences user facility. This work was funded by the Laboratory Directed Research and Development program at Sandia National Laboratories, a multimission laboratory managed and operated by National Technology and Engineering Solutions of Sandia, LLC, a wholly owned subsidiary of Honeywell International, Inc., for the U.S. Department of Energy's National Nuclear Security Administration under contract DE-NA0003525. This paper describes objective technical results and analysis. Any subjective views or opinions that might be expressed in the paper do not necessarily represent the views of the U.S. Department of Energy or the United States Government.

**REFERENCES:**

[1] M. C. Lemme, D. Akinwande, C. Huyghebaert, and C. Stampfer, "2D materials for future heterogeneous electronics," *Nature communications,* vol. 13, p. 1392, 2022.

[2] F. Zhou and Y. Chai, "Near-sensor and in-sensor computing," *Nature Electronics,* vol. 3, pp. 664-671, 2020.

[3] F. Liao, F. Zhou, and Y. Chai, "Neuromorphic vision sensors: Principle, progress and perspectives," *Journal of Semiconductors,* vol. 42, p. 013105, 2021.

[4] C. Choi, H. Seung, and D.-H. Kim, "Bio-inspired electronic eyes and synaptic photodetectors for mobile artificial vision," *IEEE Journal on Flexible Electronics,* vol. 1, pp. 76-87, 2022.

[5] J. Bian, Z. Cao, and P. Zhou, "Neuromorphic computing: Devices, hardware, and system application facilitated by two-dimensional materials," *Applied Physics Reviews,* vol. 8, 2021.

[6] L. Mennel, J. Symonowicz, S. Wachter, D. K. Polyushkin, A. J. Molina-Mendoza, and T. Mueller, "Ultrafast machine vision with 2D material neural network image sensors," *Nature,* vol. 579, pp. 62-66, 2020.

[7] H. Jang, C. Liu, H. Hinton, M. H. Lee, H. Kim, M. Seol, *et al.*, "An atomically thin optoelectronic machine vision processor," *Advanced Materials,* vol. 32, p. 2002431, 2020.

[8] K. Liu, T. Zhang, B. Dang, L. Bao, L. Xu, C. Cheng, *et al.*, "An optoelectronic synapse based on α-In2Se3 with controllable temporal dynamics for multimode and multiscale reservoir computing," *Nature Electronics*, vol. 5, pp. 761-773, 2022.

[9] B. Huang, N. Li, Q. Wang, C. Ouyang, C. He, L. Zhang, *et al.*, "Optoelectronic Synapses Based on MoS2 Transistors for Accurate Image Recognition," *Advanced Materials Interfaces,* vol. 9, p. 2201558, 2022.

[10] Y. Liu, J. Guo, E. Zhu, L. Liao, S.-J. Lee, M. Ding, *et al.*, "Approaching the Schottky–Mott limit in van der Waals metal–semiconductor junctions," *Nature,* vol. 557, pp. 696-700, 2018.

[11] P.-C. Shen, C. Su, Y. Lin, A.-S. Chou, C.-C. Cheng, J.-H. Park, *et al.*, "Ultralow contact resistance between semimetal and monolayer semiconductors," *Nature,* vol. 593, pp. 211-217, 2021.

[12] W. Li, X. Gong, Z. Yu, L. Ma, W. Sun, S. Gao, *et al.*, "Approaching the quantum limit in two-dimensional semiconductor contacts," *Nature,* vol. 613, pp. 274-279, 2023.
[13] Y. Liu, P. Stradins, and S.-H. Wei, "Van der Waals metal-semiconductor junction: Weak Fermi level pinning enables effective tuning of Schottky barrier," *Science advances,* vol. 2, p. e1600069, 2016.
[14] G.-S. Kim, S.-H. Kim, J. Park, K. H. Han, J. Kim, and H.-Y. Yu, "Schottky barrier height engineering for electrical contacts of multilayered MoS2 transistors with reduction of metal-induced gap states," *ACS nano,* vol. 12, pp. 6292-6300, 2018.
[15] H. Liu, A. T. Neal, and P. D. Ye, "Channel length scaling of MoS2 MOSFETs," *ACS nano,* vol. 6, pp. 8563-8569, 2012.
[16] J. Kang, W. Liu, and K. Banerjee, "High-performance MoS2 transistors with low-resistance molybdenum contacts," *Applied Physics Letters,* vol. 104, p. 093106, 2014.
[17] C. D. English, G. Shine, V. E. Dorgan, K. C. Saraswat, and E. Pop, "Improved contacts to MoS2 transistors by ultra-high vacuum metal deposition," *Nano letters,* vol. 16, pp. 3824-3830, 2016.
[18] A. Sebastian, R. Pendurthi, T. H. Choudhury, J. M. Redwing, and S. Das, "Benchmarking monolayer MoS2 and WS2 field-effect transistors," *Nature communications,* vol. 12, p. 693, 2021.
[19] S. Seo, S.-H. Jo, S. Kim, J. Shim, S. Oh, J.-H. Kim, *et al.*, "Artificial optic-neural synapse for colored and color-mixed pattern recognition," *Nature Communications,* vol. 9, p. 5106, 2018.
[20] J. Tao, J. S. Vazquez, H. U. Chae, R. Ahsan, and R. Kapadia, "Machine Vision With InP Based Floating-Gate Photo-Field-Effective Transistors for Color-Mixed Image Recognition," *IEEE Journal of Quantum Electronics,* vol. 58, pp. 1-7, 2022.
[21] S. Mukherjee, D. Dutta, M. Uzhansky, and E. Koren, "Monolithic In2Se3–In2O3 heterojunction for multibit non-volatile memory and logic operations using optoelectronic inputs," *npj 2D Materials and Applications,* vol. 6, p. 37, 2022.
[22] X. Fu, T. Li, B. Cai, J. Miao, G. N. Panin, X. Ma, *et al.*, "Graphene/MoS2−xOx/graphene photomemristor with tunable non-volatile responsivities for neuromorphic vision processing," *Light: Science & Applications,* vol. 12, p. 39, 2023.
[23] D. Li, B. Wang, M. Chen, J. Zhou, and Z. Zhang, "Gate-Controlled BP–WSe2 Heterojunction Diode for Logic Rectifiers and Logic Optoelectronics," *Small,* vol. 13, p. 1603726, 2017.
[24] W. Zhen, X. Zhou, S. Weng, W. Zhu, and C. Zhang, "Ultrasensitive, ultrafast, and gate-tunable two-dimensional photodetectors in ternary rhombohedral ZnIn2S4 for optical neural networks," *ACS Applied Materials & Interfaces,* vol. 14, pp. 12571-12582, 2022.
[25] C. C. S. Maria, R. A. Patil, D. P. Hasibuan, C. S. Saragih, C.-C. Lai, Y. Liou, *et al.*, "White-light photodetection enhancement and thin film impediment in Bi2S3 nanorods/thin-films homojunction photodetectors," *Applied Surface Science,* vol. 584, p. 152608, 2022.
[26] W. Zhou, Y. Zhou, Y. Peng, Y. Zhang, Y. Yin, and D. Tang, "Ultrahigh sensitivity and gain white light photodetector based on GaTe/Sn: CdS nanoflake/nanowire heterostructures," *Nanotechnology,* vol. 25, p. 445202, 2014.
[27] J. S. Rana, S. Das, and S. Jit, "High Responsive Al/PTB7/Si/Al Vertical Structure Based White Light Photodetector using FTM Method," *IEEE Photonics Technology Letters,* 2023.
[28] S. M. Yadav and A. Pandey, "An efficient white-light photodetector based on 2D-SnS 2 nanosheets," *IEEE Transactions on Electron Devices,* vol. 69, pp. 1889-1893, 2022.

[29] Z. Yang, L. Liao, F. Gong, F. Wang, Z. Wang, X. Liu, *et al.*, "WSe2/GeSe heterojunction photodiode with giant gate tunability," *Nano Energy,* vol. 49, pp. 103-108, 2018.

[30] D. Li, M. Chen, Z. Sun, P. Yu, Z. Liu, P. M. Ajayan, *et al.*, "Two-dimensional non-volatile programmable p–n junctions," *Nature nanotechnology,* vol. 12, pp. 901-906, 2017.

[31] B. W. Baugher, H. O. Churchill, Y. Yang, and P. Jarillo-Herrero, "Optoelectronic devices based on electrically tunable p–n diodes in a monolayer dichalcogenide," *Nature nanotechnology,* vol. 9, pp. 262-267, 2014.

[32] M. Dhakshnamoorthy, A. Kathirvel, S. M. Raj, V. R. Ancha, M. Abebe, and S. K. Batabyal, "Self-powered white light photodetector with enhanced photoresponse using camphor sulphonic acid treated CsPbBr3 perovskite in carbon matrix," *Materials Letters,* vol. 341, p. 134250, 2023.

[33] A. J. Molina-Mendoza, M. Paur, and T. Mueller, "Nonvolatile programmable WSe2 photodetector," *Advanced Optical Materials,* vol. 8, p. 2000417, 2020.

[34] A. K. Rana, M. Patel, T. T. Nguyen, J.-H. Yun, and J. Kim, "Transparent Co3O4/ZnO photovoltaic broadband photodetector," *Materials Science in Semiconductor Processing,* vol. 117, p. 105192, 2020.

[35] A. Kathirvel, A. U. Maheswari, and M. Sivakumar, "Highly sensitive and wavelength-tunable solution-processed BiFeO3 heterojunction based fast-response self-powered white-light photodetector," *Thin Solid Films,* vol. 761, p. 139534, 2022.

[36] C.-Y. Wang, S.-J. Liang, S. Wang, P. Wang, Z. a. Li, Z. Wang, *et al.*, "Gate-tunable van der Waals heterostructure for reconfigurable neural network vision sensor," *Science Advances,* vol. 6, p. eaba6173, 2020.

[37] Y. Guo, C. Liu, H. Tanaka, and E. Nakamura, "Air-stable and solution-processable perovskite photodetectors for solar-blind UV and visible light," *The journal of physical chemistry letters,* vol. 6, pp. 535-539, 2015.

[38] H. Fang, Q. Li, J. Ding, N. Li, H. Tian, L. Zhang, *et al.*, "A self-powered organolead halide perovskite single crystal photodetector driven by a DVD-based triboelectric nanogenerator," *Journal of Materials Chemistry C,* vol. 4, pp. 630-636, 2016.

[39] A. Quellmalz, X. Wang, S. Sawallich, B. Uzlu, M. Otto, S. Wagner, *et al.*, "Large-area integration of two-dimensional materials and their heterostructures by wafer bonding," *Nature communications,* vol. 12, p. 917, 2021.

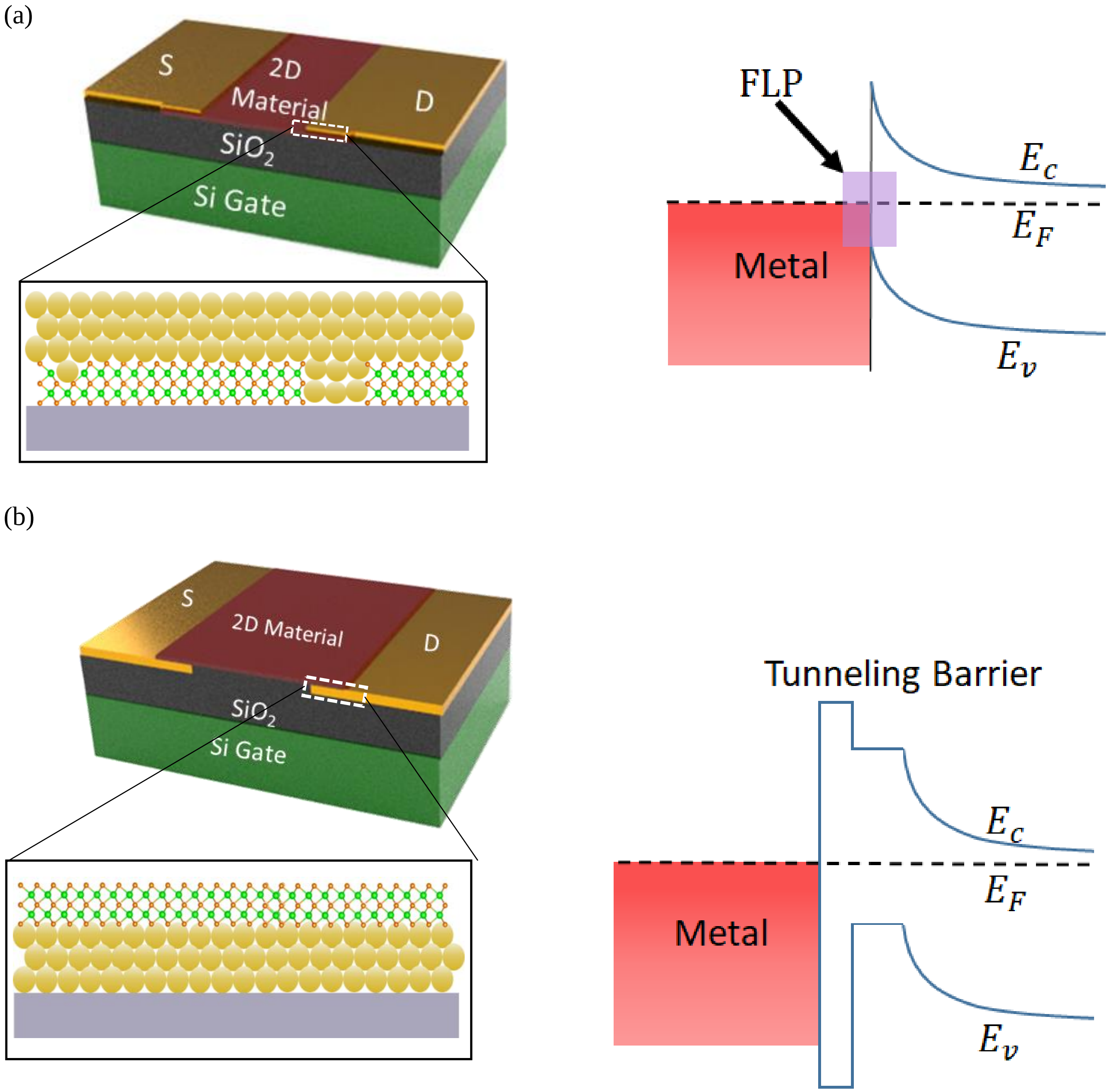

**Figure 1**: **Semiconductor-Metal Contact**: structure schematics and energy band diagram of (a) conventional 2D-FET device fabricated via metal evaporation on top of 2D material. The energy band diagram shows Fermi Level Pinning (FLP) at the contacts (b) vdW-FET structure with clean interface between 2D material and metal contact.

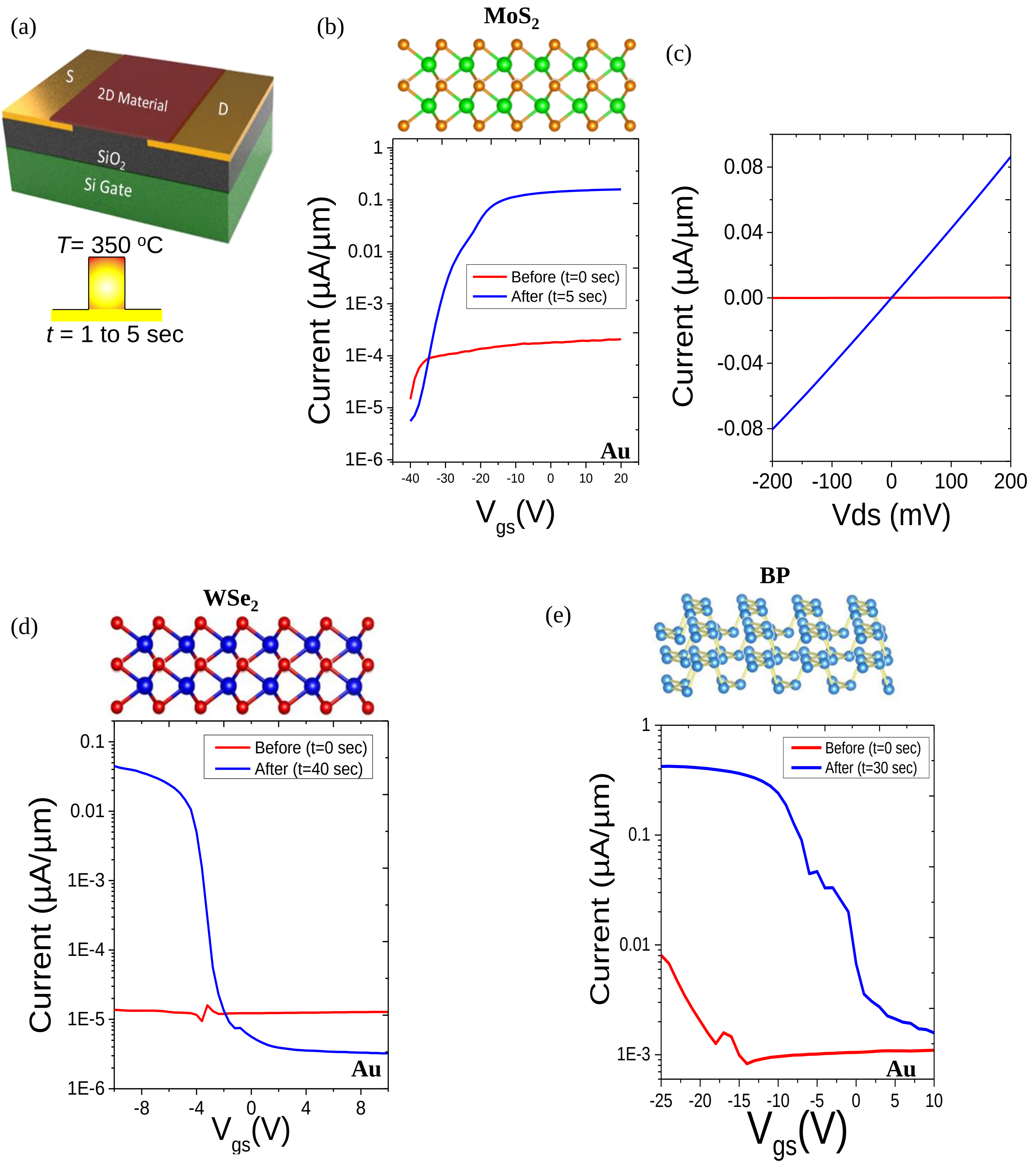

**Figure 2. Performance of vdW-FETs.** (a) anneal and heal process showing thermal pulse applied to vdW-FET device. (b) $IV_{ds}$ and (c) $IV_{gs}$ response of $MoS_2$ vdW-FET before and after anneal and heal process. The channel length is $L$=2um. $IV_{gs}$ response before and after anneal and heal treatment process for (d) $WSe_2$ and (e) Layered black phosphorus.

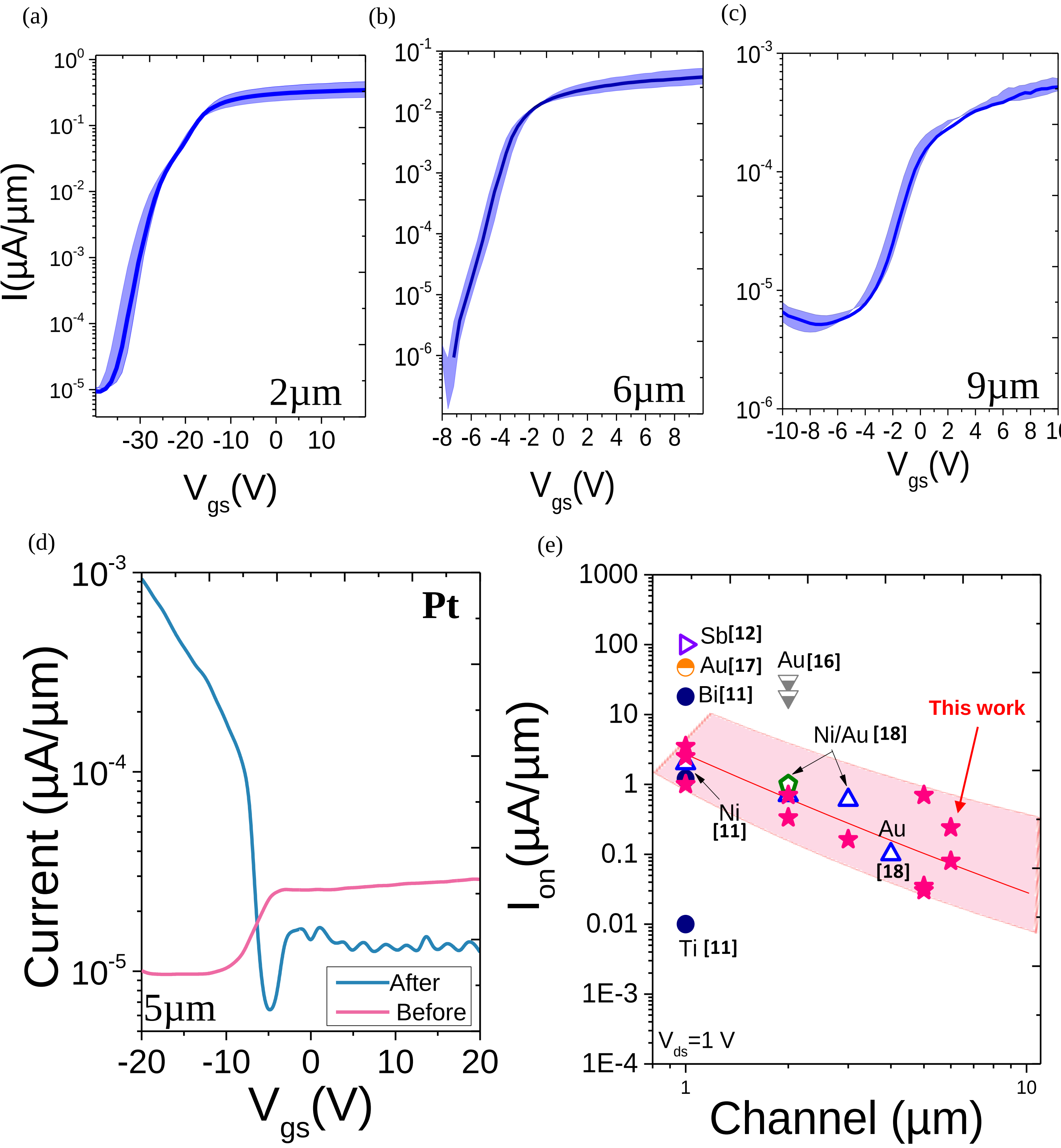

**Figure 3. Electrical Characteristics of of $MoS_2$ vdW-FETs.** (a) $IV_{gs}$ measurements showing the highest current obtained for $MoS_2$ devices with (a) $L$=2μm (b) $L$=6μm and (c) $L$=9μm. The maximum $I_{on}$ decreases with increasing channel length. (d) $IV_{gs}$ of $MoS_2$ device with Pt electrodes showing *p*-type behavior. (e) benchmarking Imax vs. channel length of different $MoS_2$ vdW-FET devices using Au with other high performance 2D-FET devices.

(a)

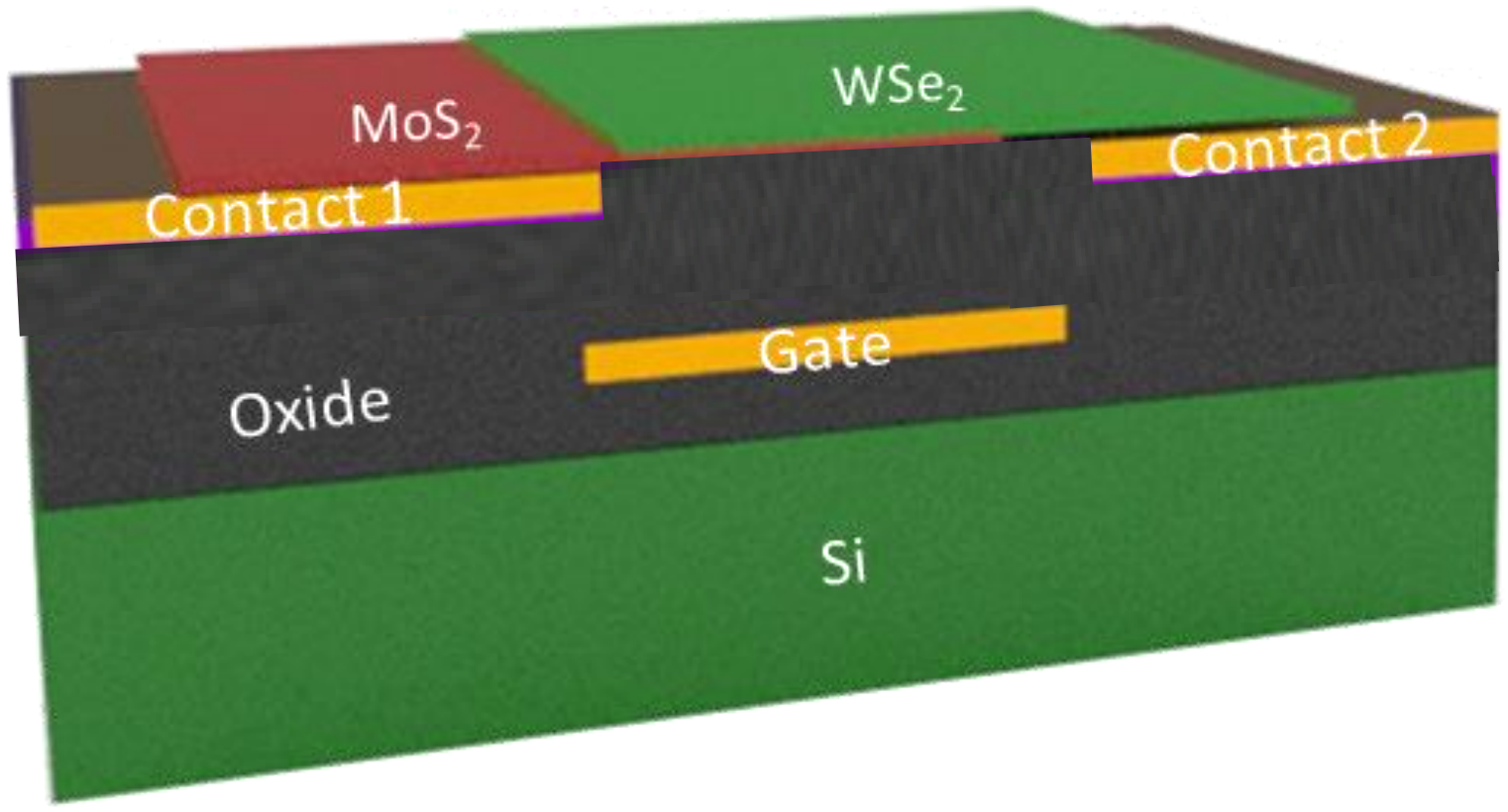

(b)

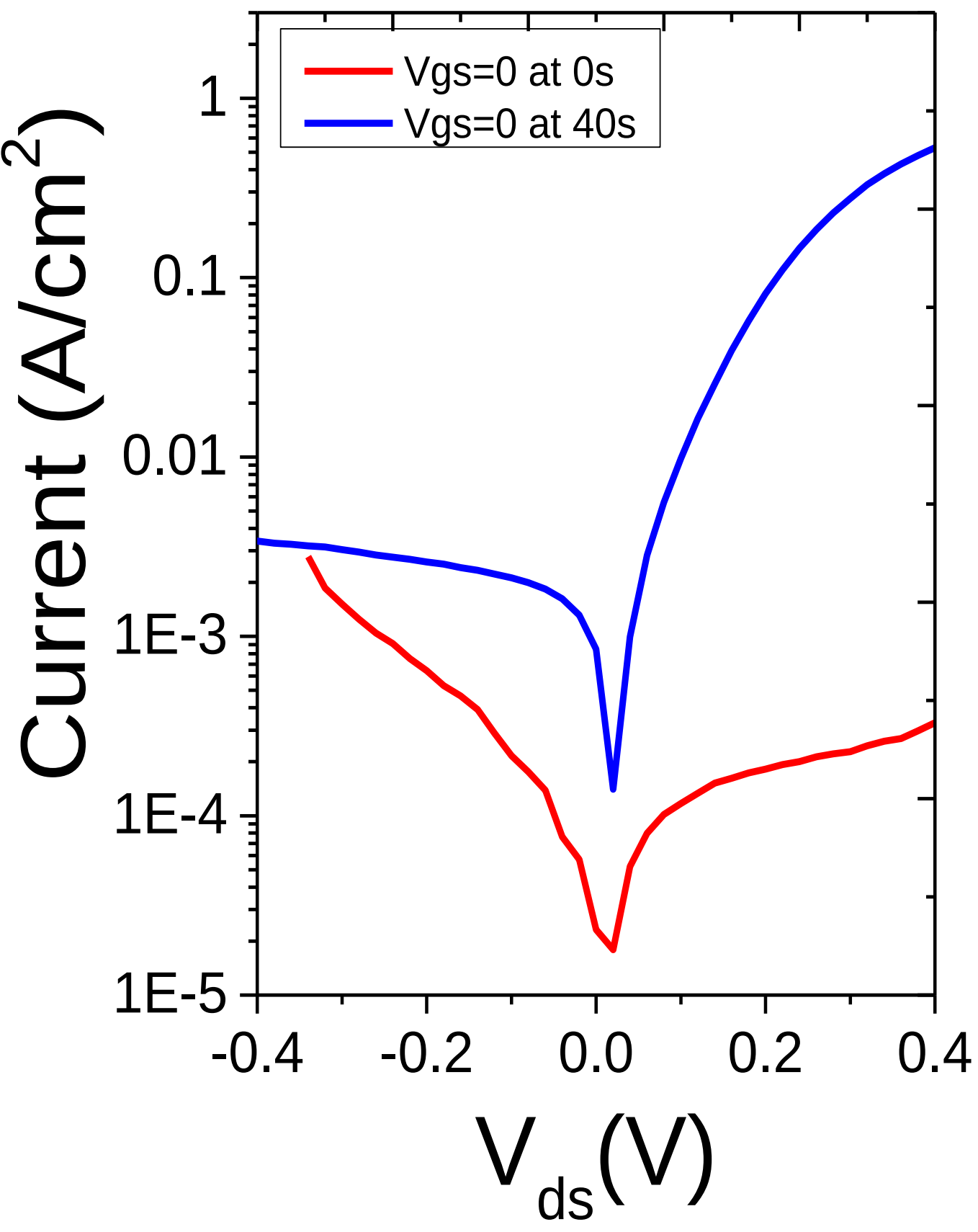

(c)

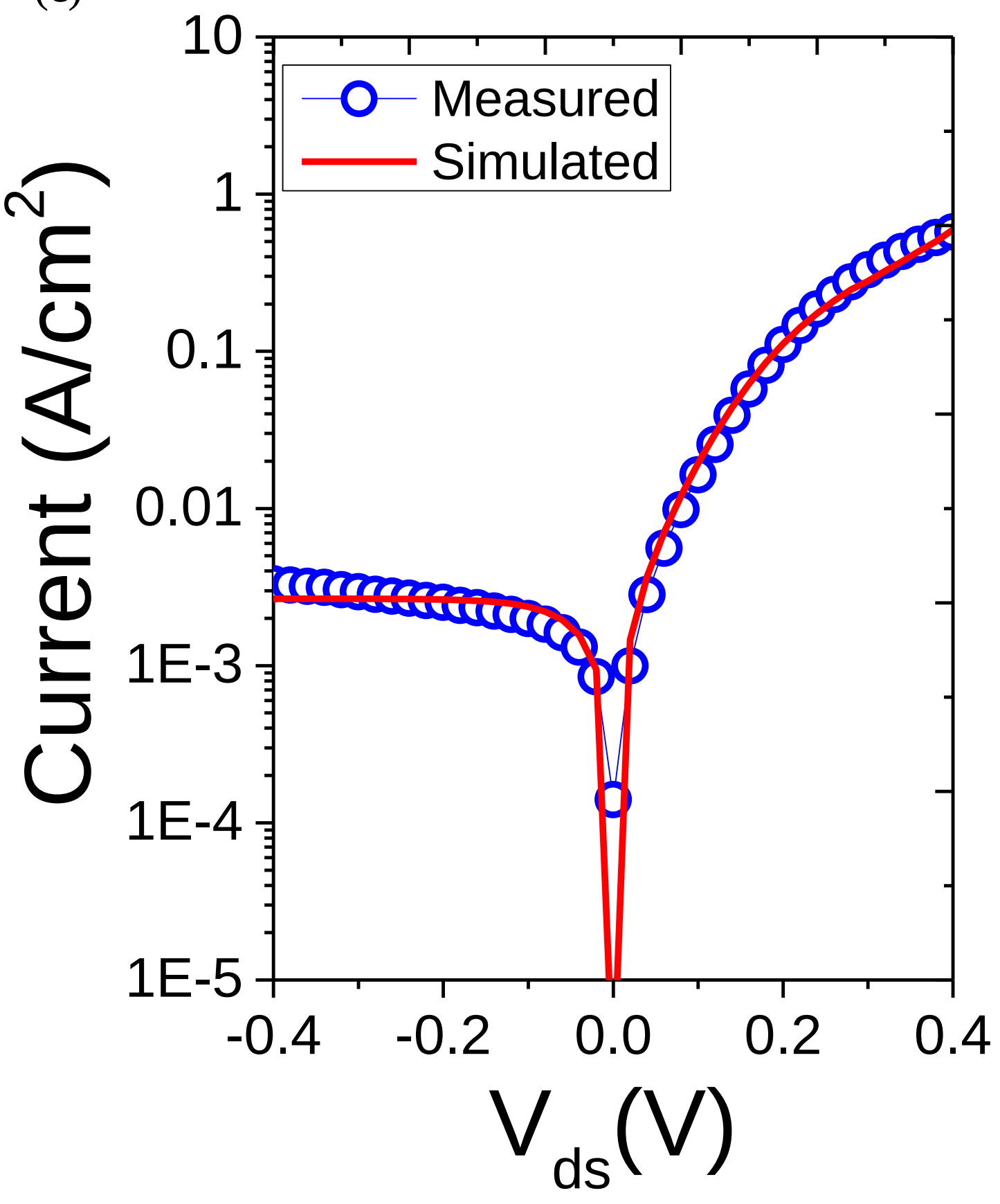

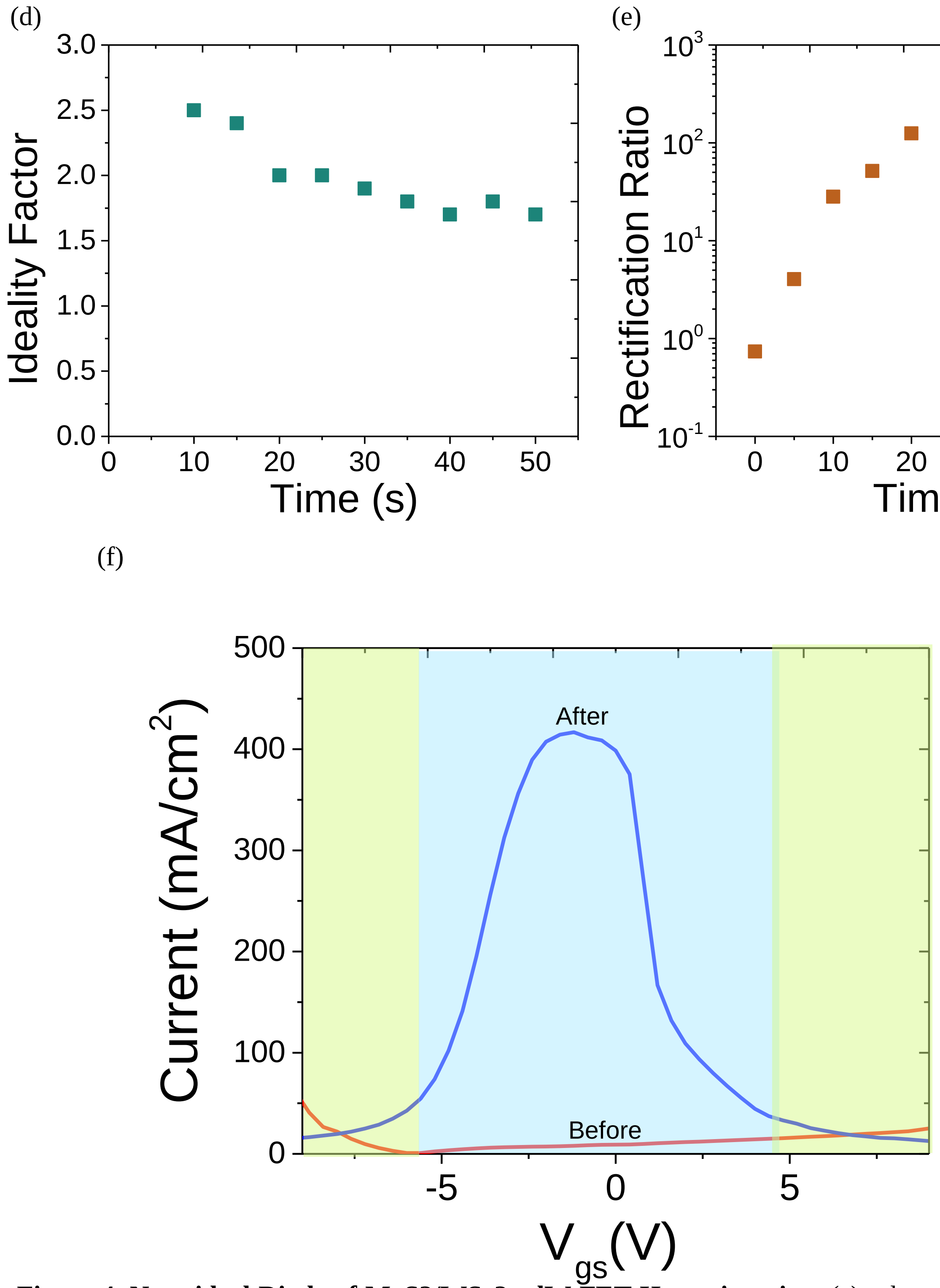

**Figure 4. Near-ideal Diode of MoS2/WSe2 vdW-FET Heterojunction.** (a) schematic diagram showing WSe2/MoS2 junction in vdW-FET configuration. (b) $IV_{ds}$ before and after anneal and heal treatment process. (c) Measured and simulated $IV_{ds}$ using ideal diode equation. The ideality factor here is ~1.65. Evolution of (d) ideality factor and (e) rectification ratio with increasing anneal and heal treatment time. (f) $IV_{gs}$ measured before (red curve) and after (blue curve) anneal and heal process showing switching between high resistive region (green shaded area) and low resistive region (blue shaded area).

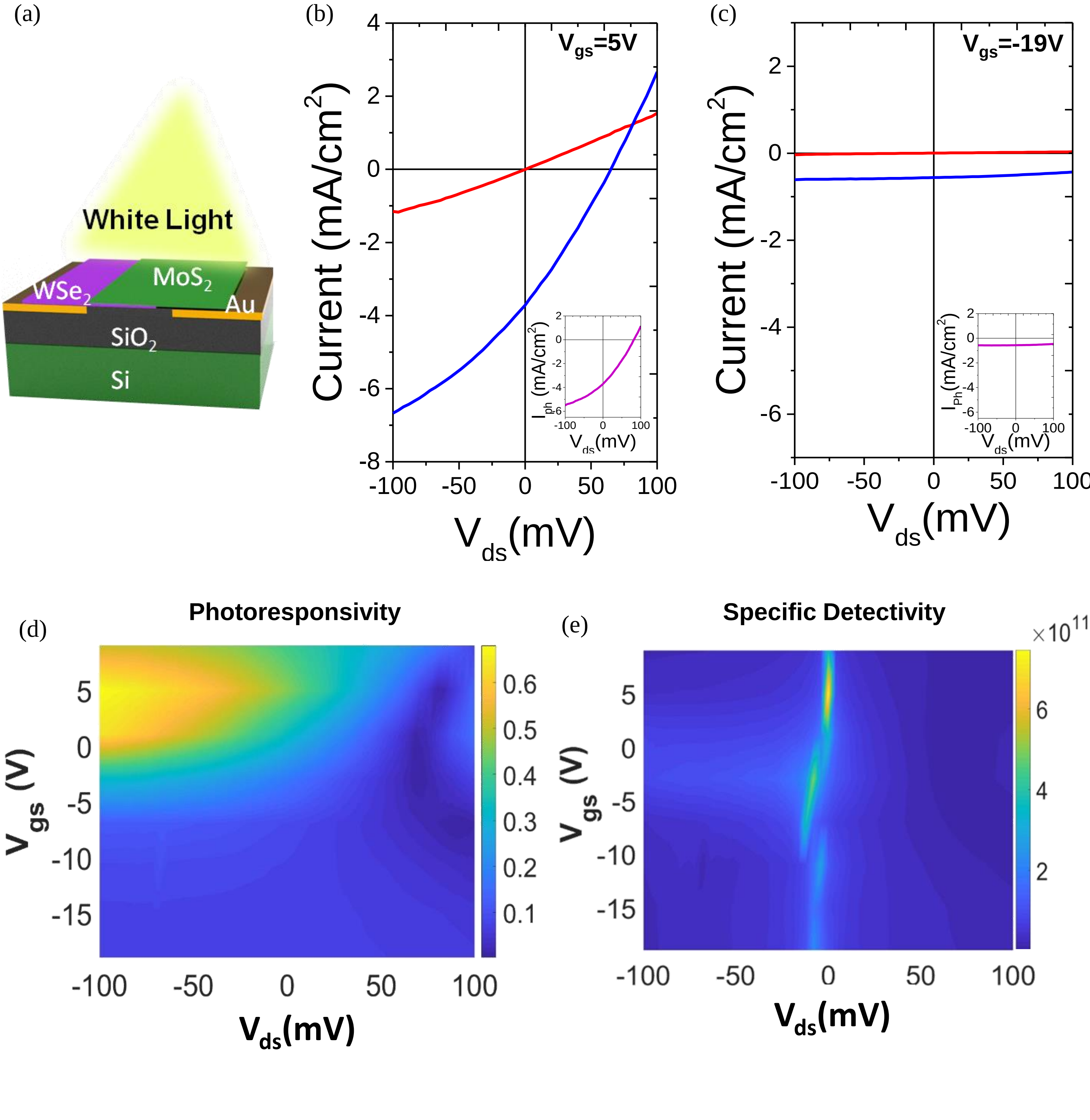

**Figure 5. Gate Tunable of $MoS_2/WSe_2$ White Light Photodetector.** (a) Schematics of vdW-FET phototransistor under white light illumination. $IV_{ds}$ measurements in dark and under white light illumination for of $MoS_2/WSe_2$ vdW-FET device after modified vdW deposition lithography under (c) $V_{gs}$=5V and (d) $V_{gs}$=-19V. Inset shows the photocurrent ($I_{ph}$) vs. applied $V_{ds}$. (e) Responsivity and (f) specific detectivity of the optical sensor at different applied $V_{gs}$ and $V_{ds}$.

(a)

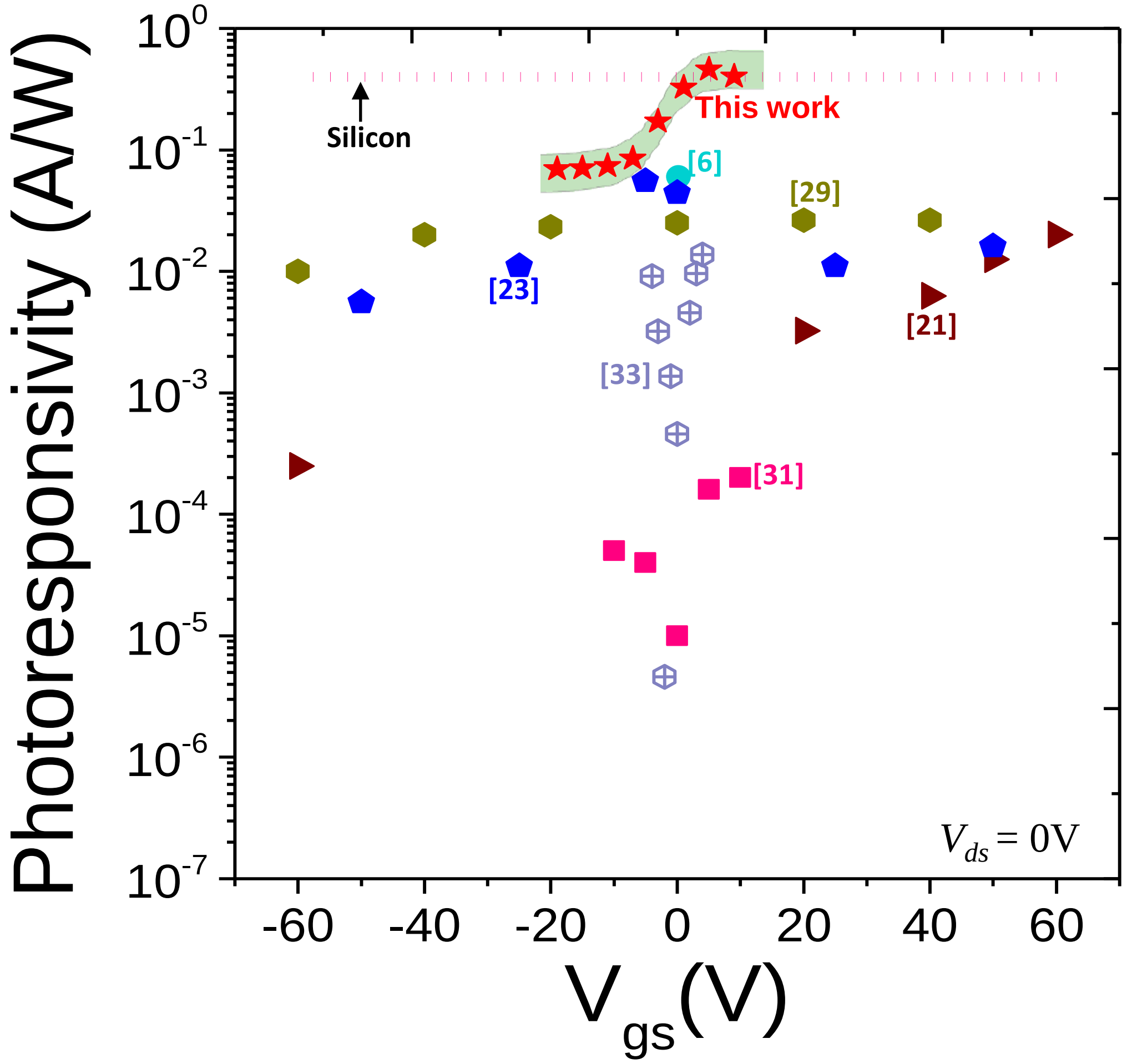

(b)

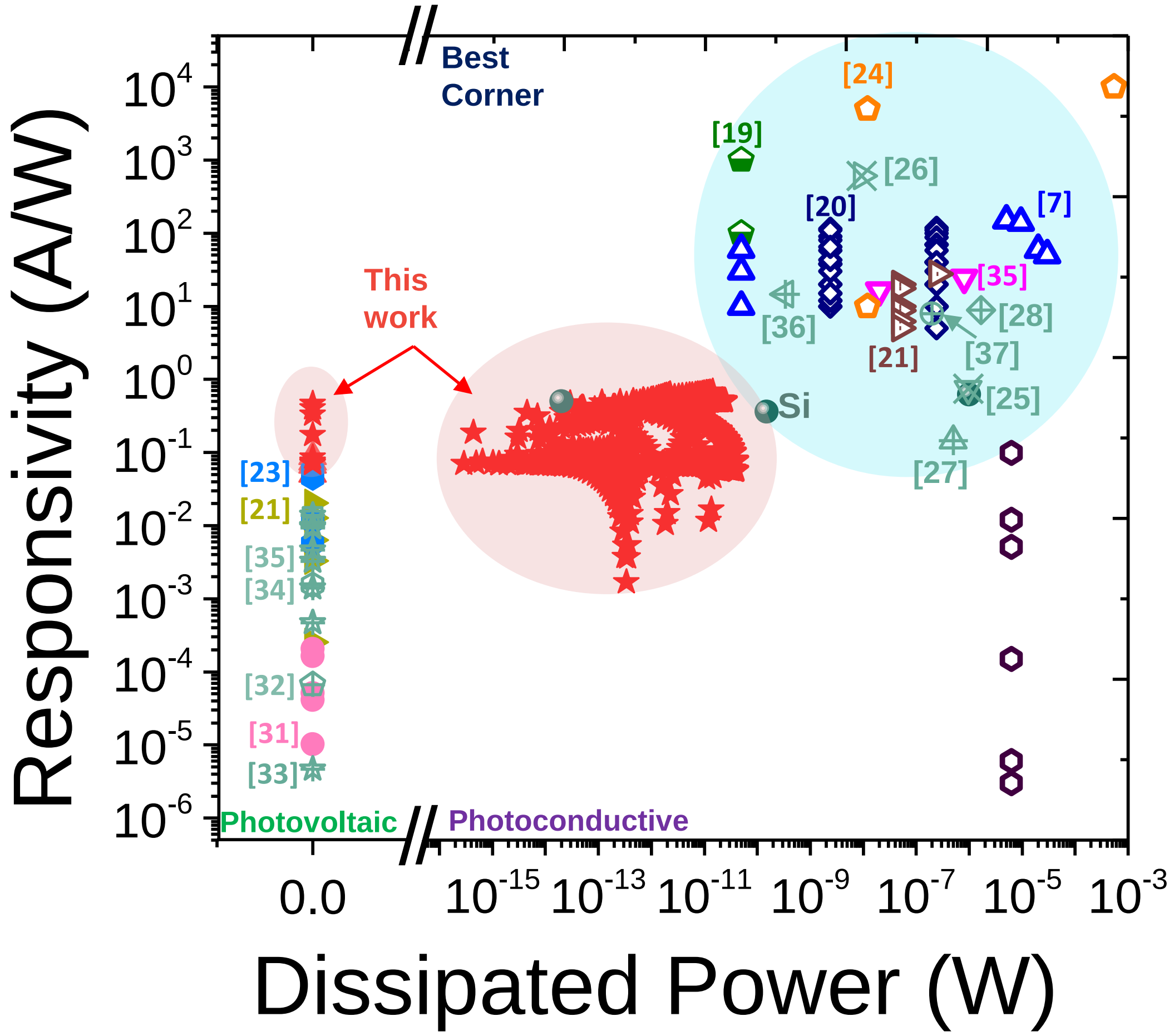

**Figure 6. Benchmark of $MoS_2/WSe_2$ vdW-FET Photodetector Technology.** (a) Responsivity vs, applied $V_{gs}$ under photovoltaic operations ($V_{ds}$ = 0V) of $MoS_2/WSe_2$ fabricated via modified vdW deposition lithography benchmarked against other machine vision technologies operating under photovoltaic mode. (b) Responsivity vs. dissipated power of gate tunable $MoS_2/WSe_2$ white light photodetector operating in photoconductive mode benchmarked against other photodetector technologies used including Si, III–Vs, 2D materials, perovskite, etc. The highlighted pink area represents our device operating region while light blue shaded area represents operating region of most reported photodetectors. The dissipated power takes into account the biasing conditions of the benchmarked photodetectors when operating under dark conditions. For both plots, filled points are photodetectors used for specific wavelength(s) (λ) detection (highlighted next to the reference in table S1). Open points with cross inside represent white light photodetectors.

# Unlocking High Performance, Ultra-Low Power Van der Waals Transistors: Towards Back-End-of-Line In-Sensor Machine Vision Applications

*Olaiyan Alolaiyan[1,†], Shahad Albwardi[1,†], Sarah Alsaggaf[1,†], Thamer Tabbakh[1], Frank W. DelRio[2], and Moh. R. Amer*[1,3,4]*

[1]Center of Excellence for Green Nanotechnologies,
Microelectronics and Semiconductor Institute
King Abdulaziz City for Science and Technology, Riyadh, 11442, Saudi Arabia
[2]Sandia National Laboratories,
Material, Physical, and Chemical Sciences Center, Albuquerque, NM, 87123, USA
[3]Department of Electrical and Computer Engineering
University of California, Los Angeles, Los Angeles, CA, 90095, USA
[4]Department of Electrical and Computer Engineering
University of Southern California, Los Angeles, CA, 90089, USA

*Please send all correspondence to mamer@seas.ucla.edu, mamer@kacst.edu.sa,
[†]These authors contributed equally

**Table S1.** Comparison of different photodetector technologies and their corresponding photodetection wavelength used in benchmarking plots in figure 6

| Ref. | Technology | Detection Wavelength |
|---|---|---|
| 6 | $WSe_2$ split gate in-sensor machine vision | 650nm |
| 7 | $MoS_2$ FET | 532nm |
| 19 | near sensor $WSe_2$ Photodetector | 405nm, 532nm, 655nm |
| 20 | InP Based Floating-Gate Photo-Field-Effective Transistors | 445nm, 532nm, 655nm |
| 21 | $In_2Se_3$–$In_2O_3$ heterojunction | 625nm and 655nm |
| 22 | Graphene/$MoS_{2-x}O_x$/graphene photomemristor | White light |
| 23 | Gate-Controlled BP–$WSe_2$ Heterojunction Diode | 640nm |
| 24 | Ternary rhombohedral $ZnIn_2S_4$ | 405nm |
| 25 | $Bi_2S_3$ nanorods/thin-films homojunction | White light |
| 26 | GaTe/Sn: CdS nanoflake/nanowire heterostructures | White light |
| 27 | Al/PTB7/Si/Al | White light |
| 28 | 2D-$SnS_2$ nanosheets | White light |
| 29 | $WSe_2$/GeSe heterojunction | 532nm |
| 31 | p–n diodes in a monolayer dichalcogenide | 700nm |
| 32 | $CsPbBr_3$ perovskite | White light |
| 33 | $WSe_2$ split gate | White light |
| 34 | Transparent $Co_3O_4$/ZnO | White light |
| 35 | solution-processed $BiFeO_3$ heterojunction | White light |
| 36 | Gate-tunable van der Waals heterostructure | 405nm, 650nm |
| 37 | solution-processable perovskite | White light |
| 38 | halide perovskite | White light |

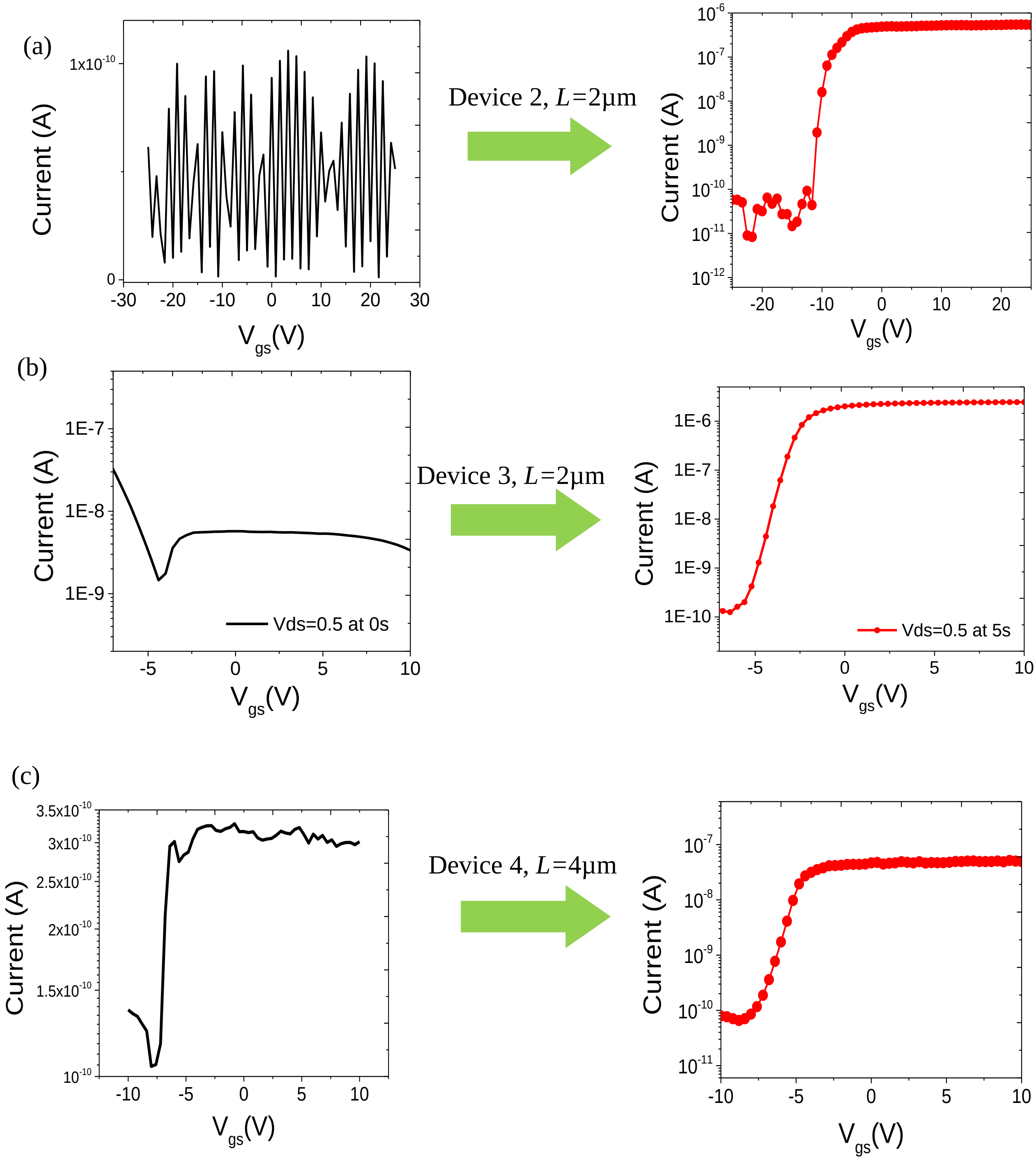

Figure S1. $IV_{ds}$ and $IV_{gs}$ measurements of 3 additional $MoS_2$ vdW-FET samples on Au electrodes showing enhanced device performance after anneal and heal treatment process. Treatment time is specified in the legned of each figure.

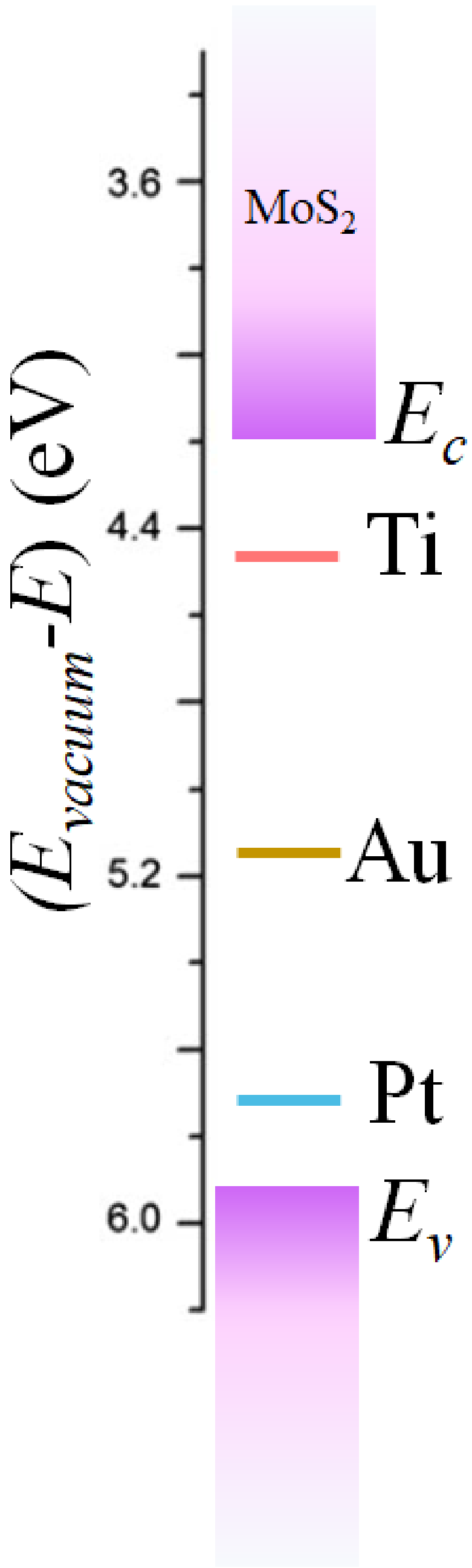

Figure S2. Energy band diagram of MoS2 showing the alignment of different metal work functions with respect to the conduction and valance band energies. Values obtained from ref. [13].

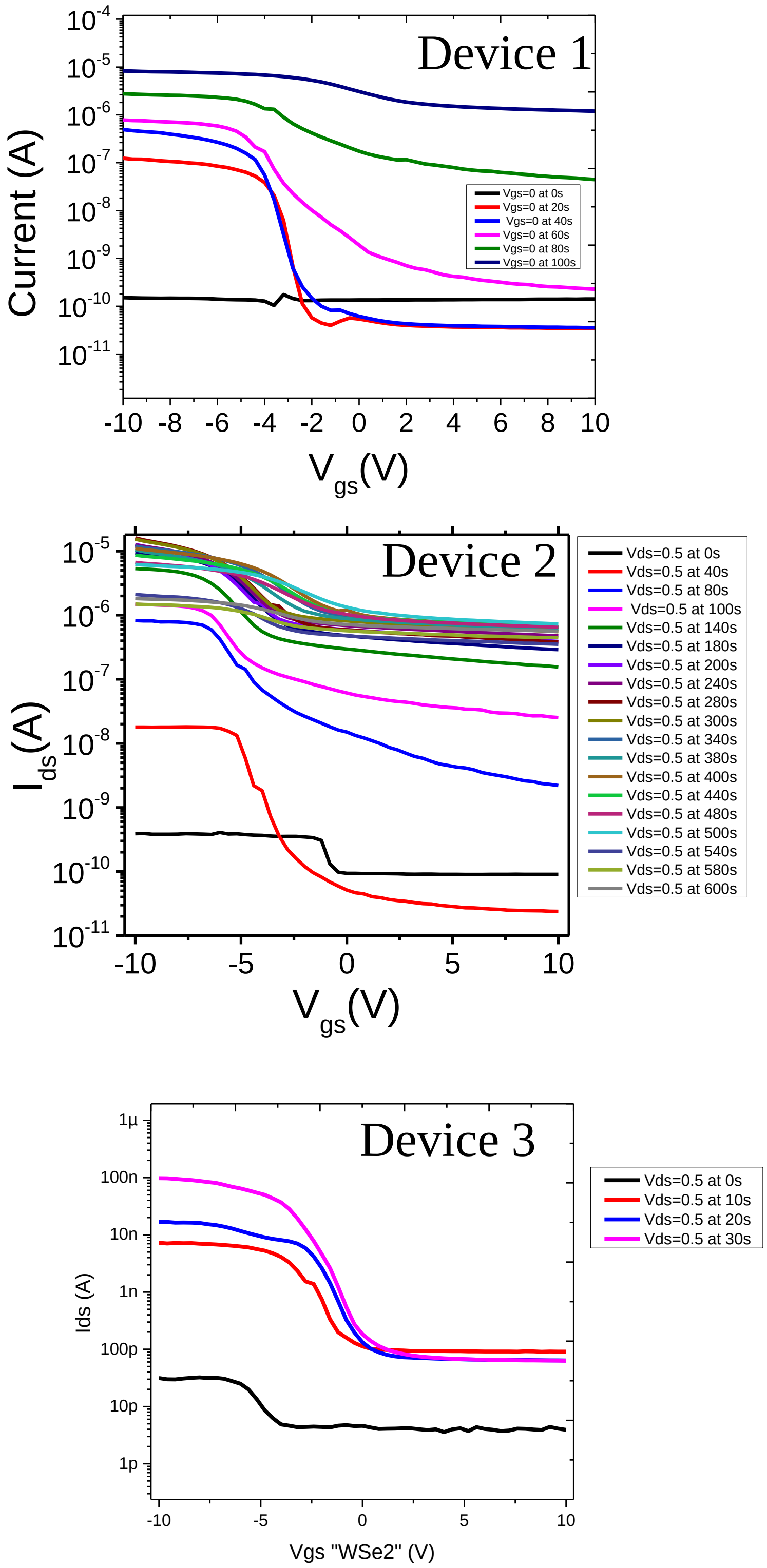

Figure S3. $IV_{gs}$ of 3 different $WSe_2$ vdW-FET devices on Au electrodes showing the evolution of device performance at different anneal and heal treatment time.

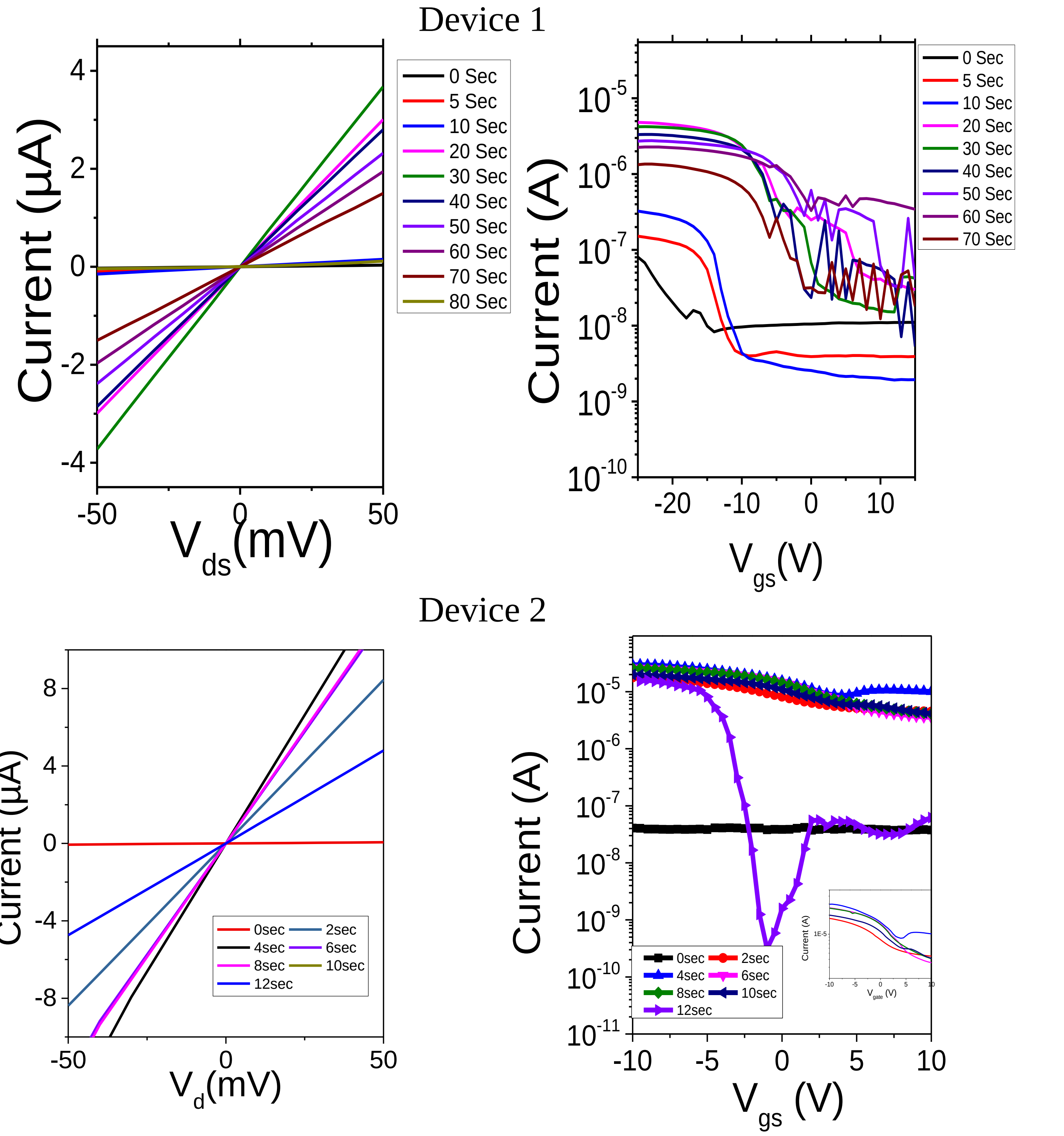

Figure S4. $IV_{ds}$ and $IV_{gs}$ measurements of BP vdW-FET devices on Au electrodes showing the evolution of device performance at different anneal and heal treatment time.

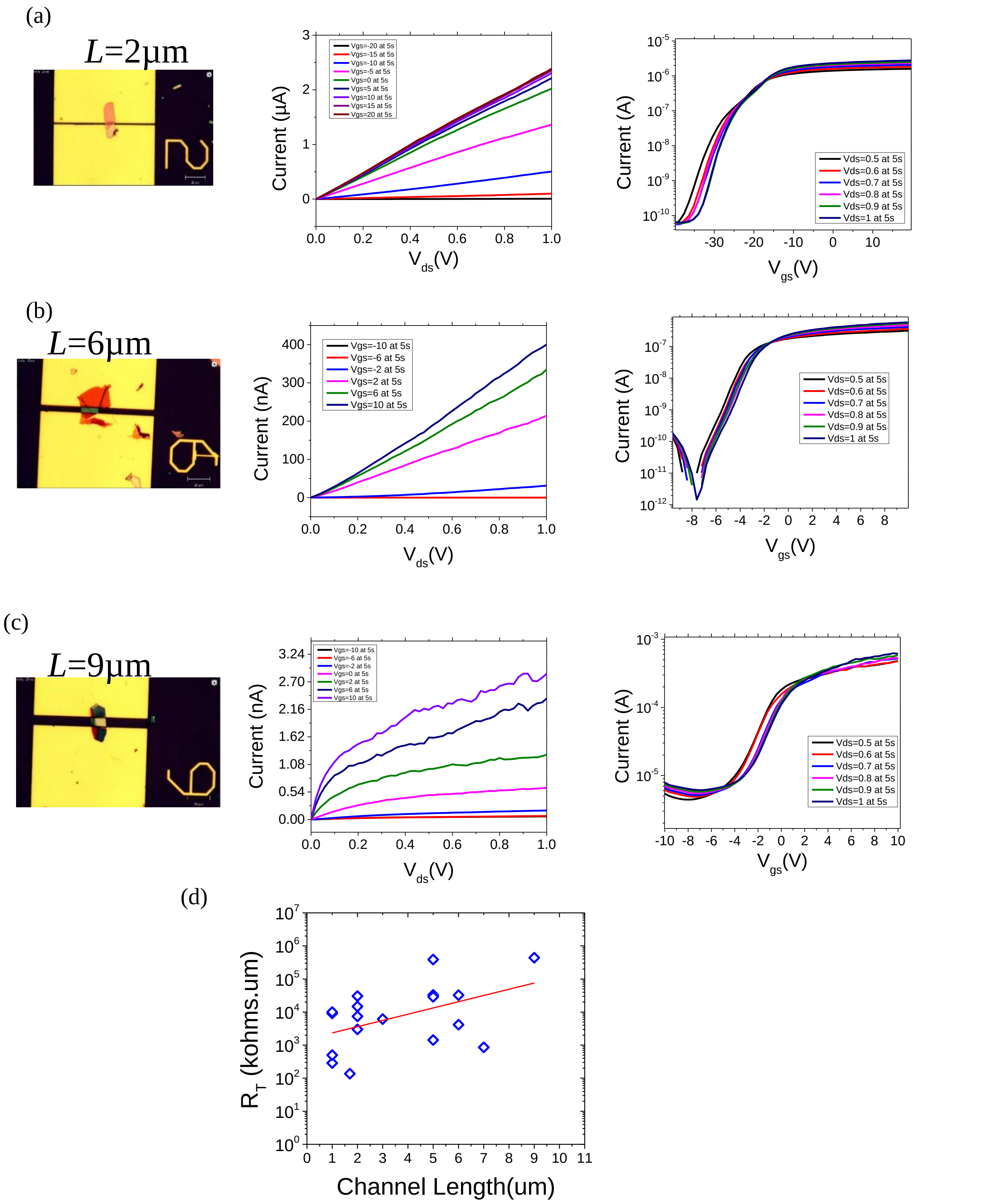

Figure S5. $IV_{ds}$ at different $V_{gs}$ and $IV_{gs}$ at different $V_{ds}$ of $MoS_2$ vdW-FET devices in figure 3 of the main manuscript with different channel lengths (a) $L$=2μm, (b) $L$=6μm, (c) $L$=9μm. (d) Device total resistance ($R_T$) vs. channel length of 17 different $MoS_2$ vdW-FET devices with Au electrodes.

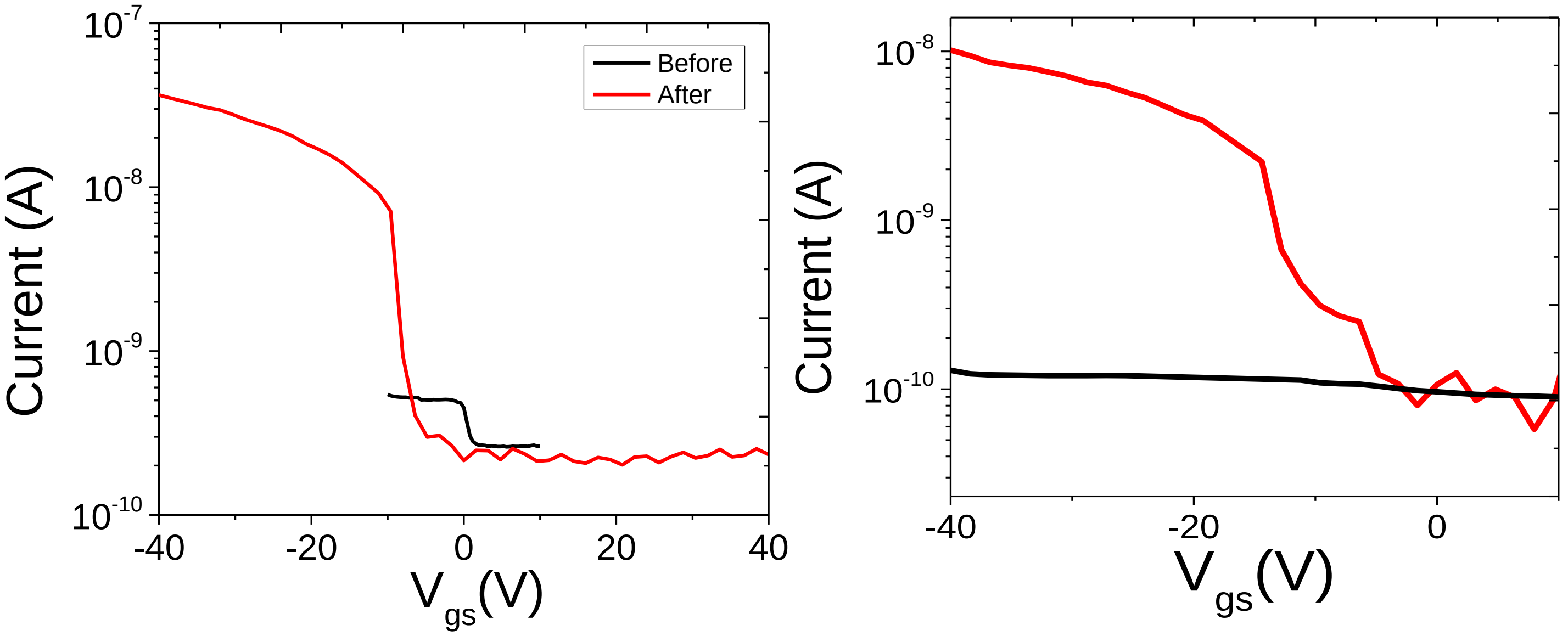

Figure S6. $IV_{gs}$ of 2 different MoS2-Pt samples showing P-type behavior in vdW-FET configuration after anneal and heal treatment.

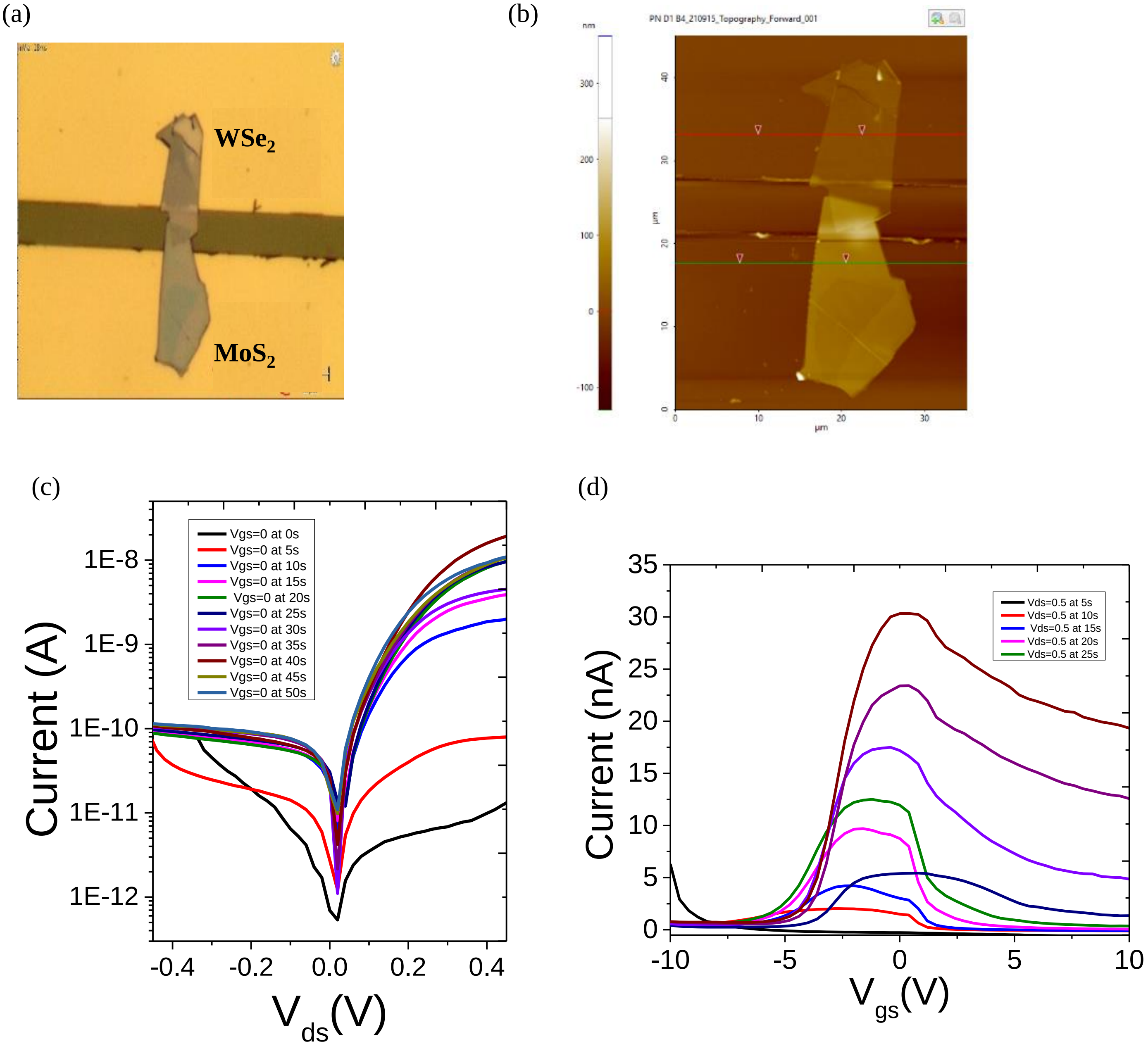

Figure S7. (a) optical image and (b) AFM image of $MoS_2/WSe_2$ junction shown in figure 4 of the manuscript. (c) $IV_{ds}$ and (d) $IV_{gs}$ after each anneal and heal treatment cycle.

(a)

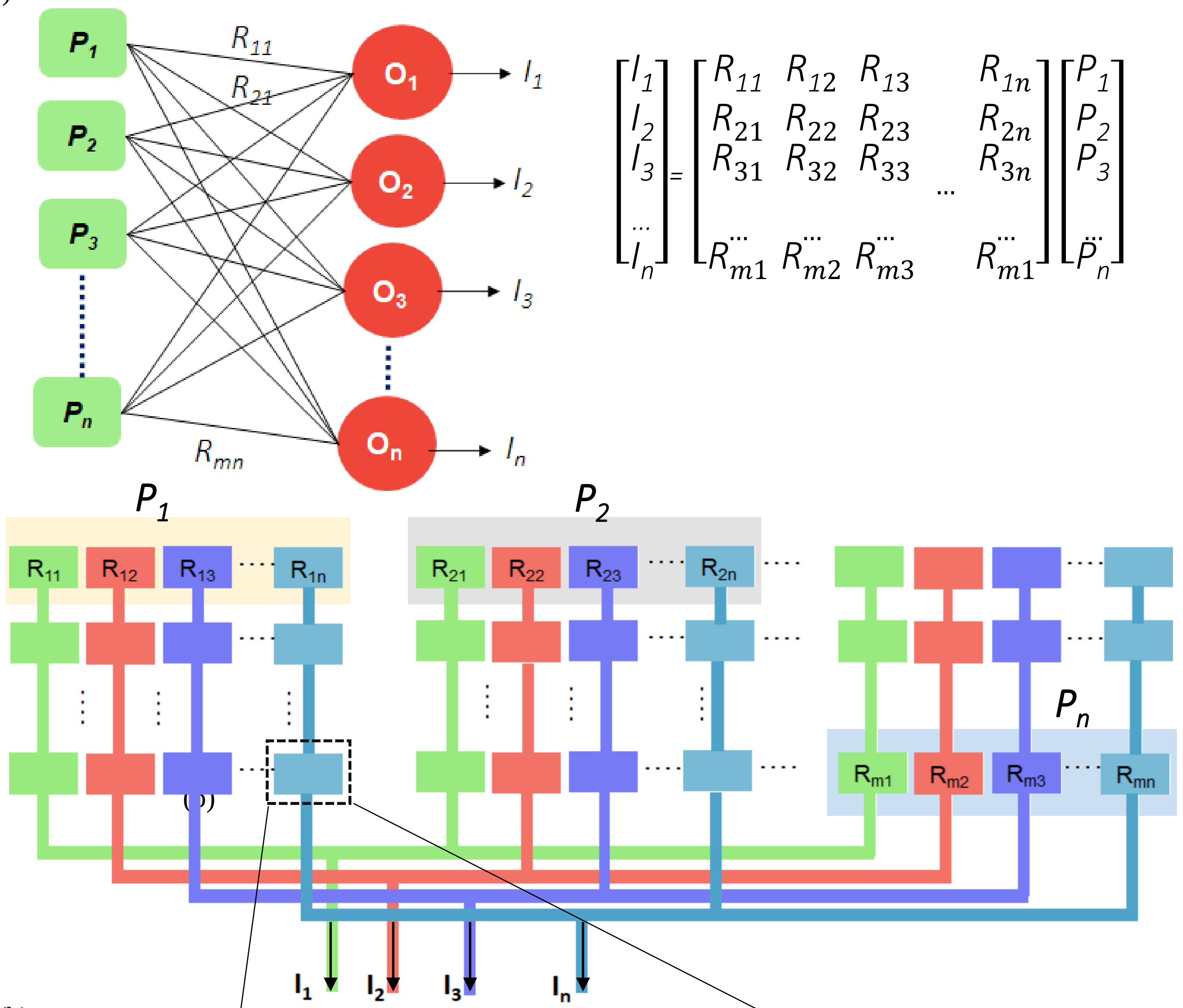

(b)

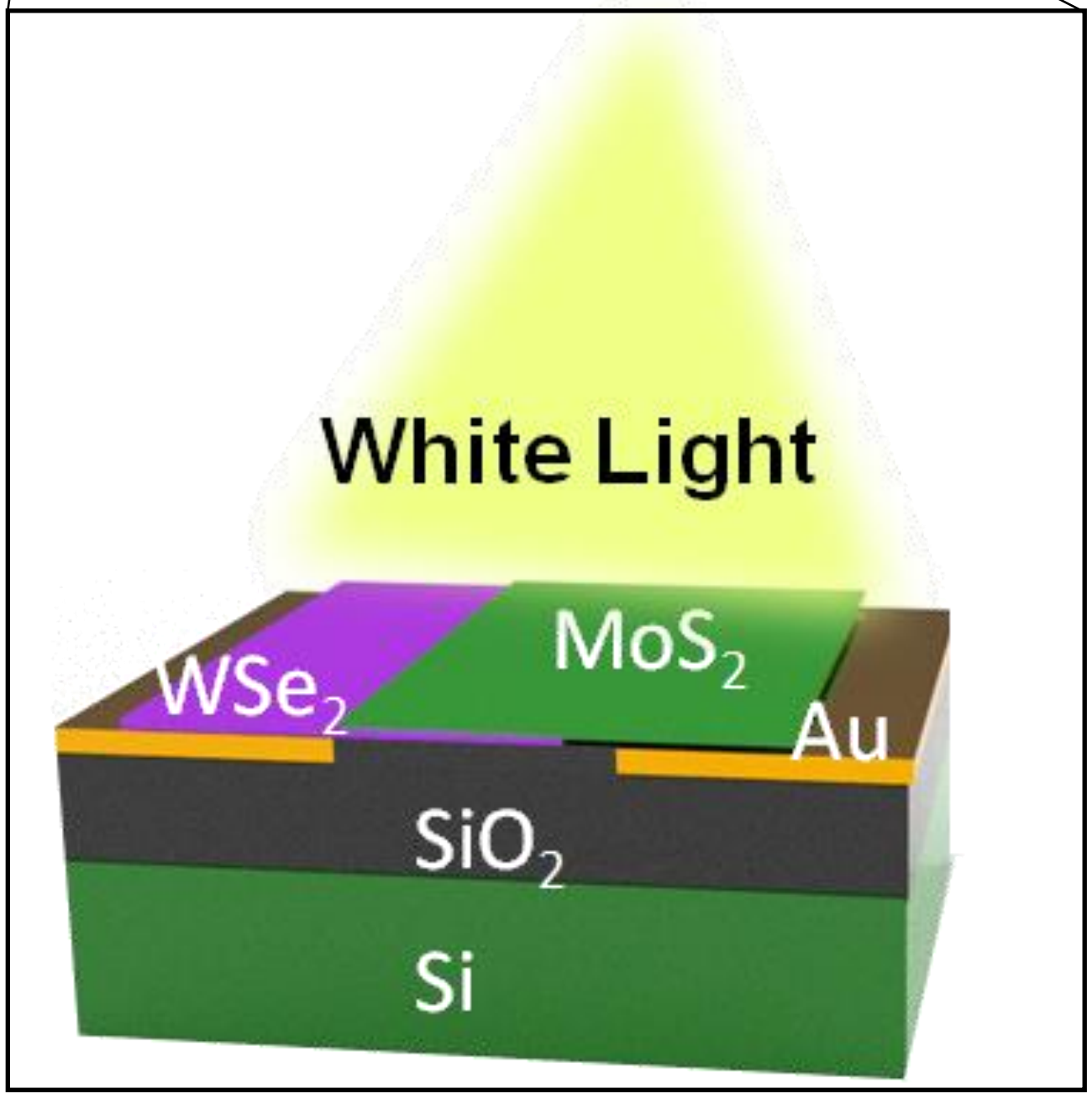

Figure S8. MAC operation concept illustration for in-sensor machine vision computing architecture using reconfigurable optical sensor. The operation is done by the matrix multiplication shown on the right side with the input being the optical power (*P*) incident on an optical sensor element with responsivity (*R*). The optical neuron (O) is the output current. (b) Schematics of proposed vdW-FET reconfigurable optical sensor with tunable responsivity.

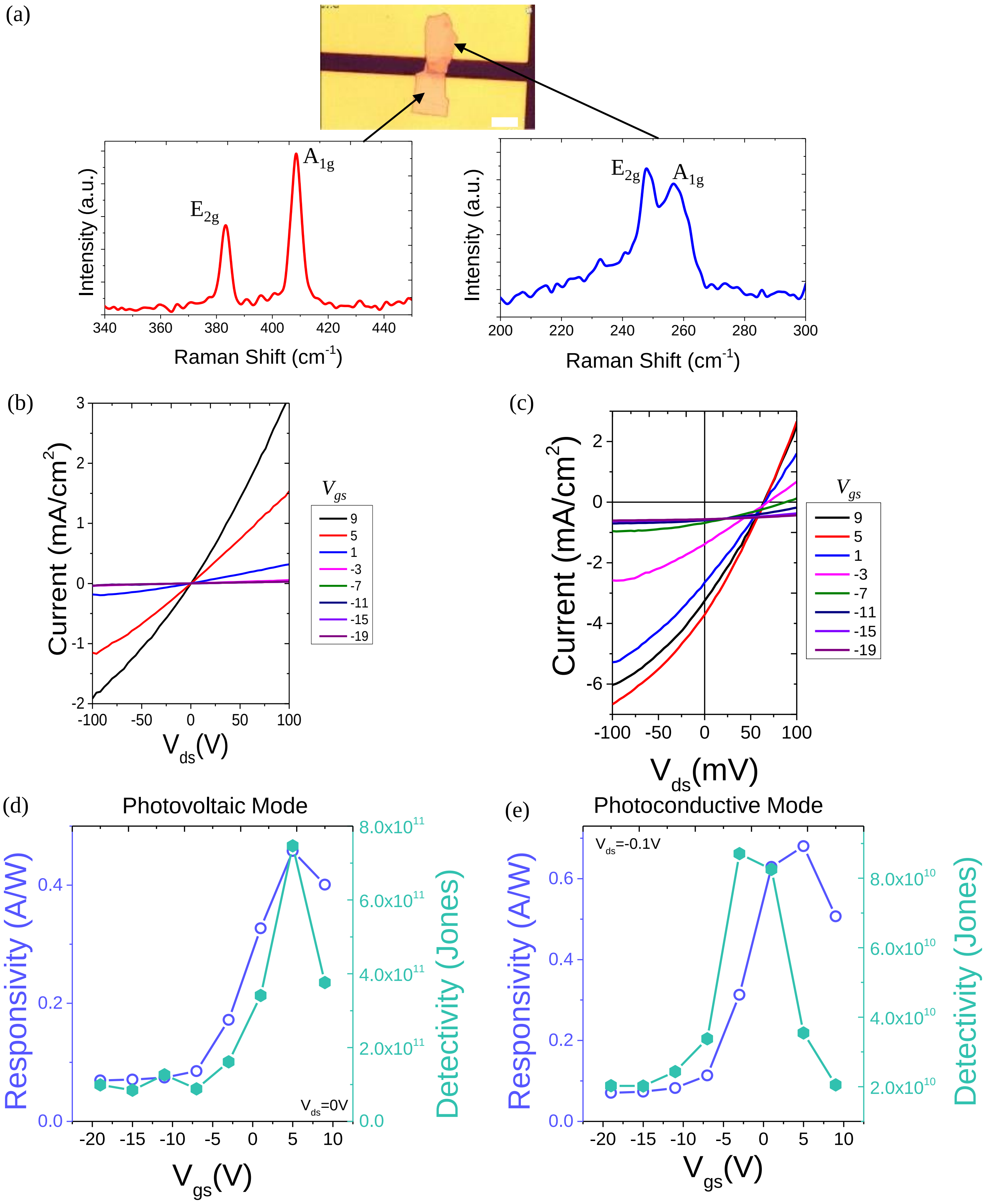

Figure S9. (a) optical image of $MoS_2/WSe_2$ photodetector fabricated on Au electrodes. The arrows point to the location of the Raman spectrum collected at that point showing Raman peaks signatures for each material. $IV_{ds}$ sampled at different applied $V_{gs}$ under (b) dark and (c) white light illumination. Responsivity and detectivity at different applied $V_{gs}$ is shown for (d) photovoltaic mode and (e) photoconductive mode.

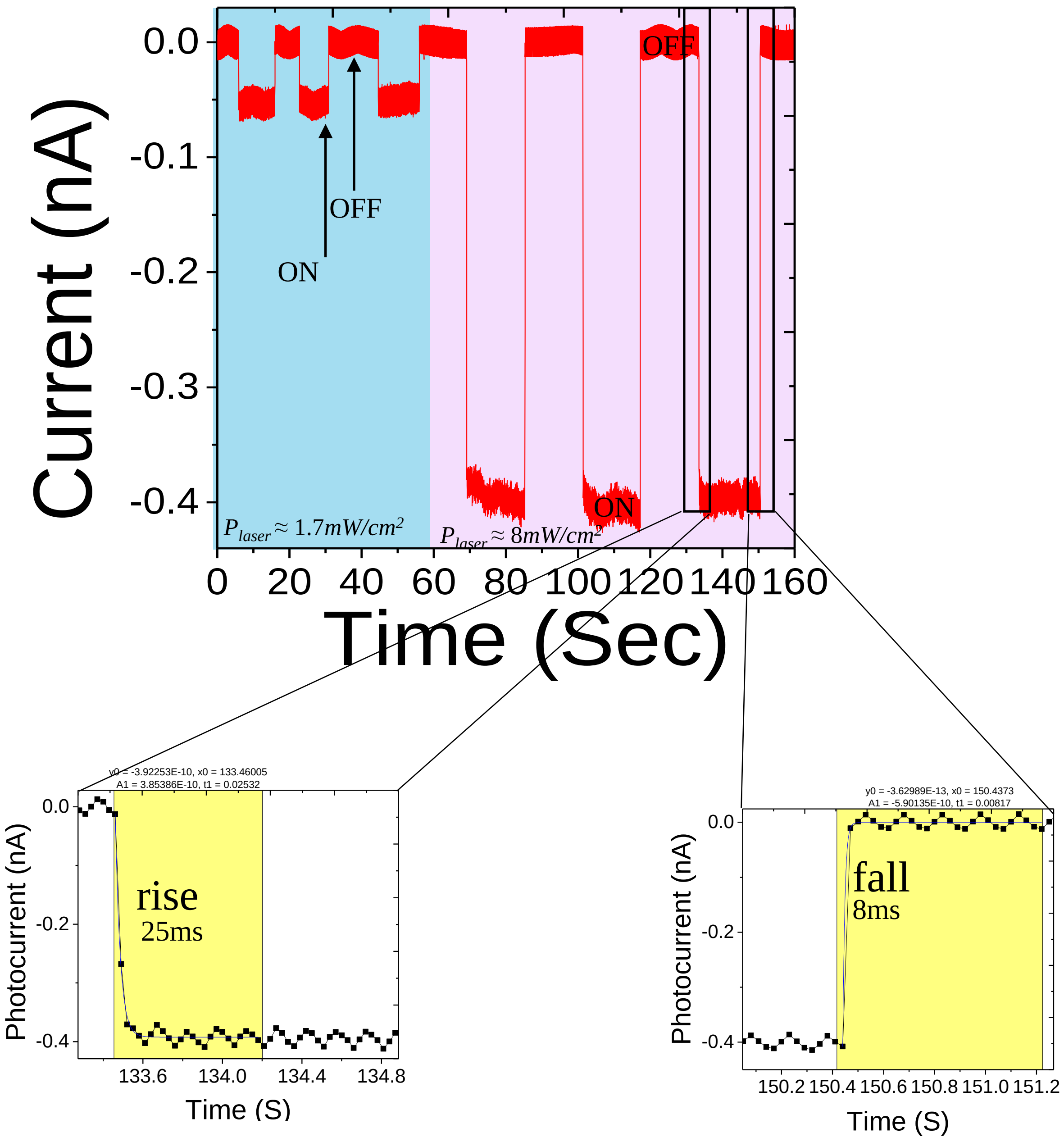

Figure S10. Photodetector transient response time showing rise and fall time of 25ms and 8ms, respectively.

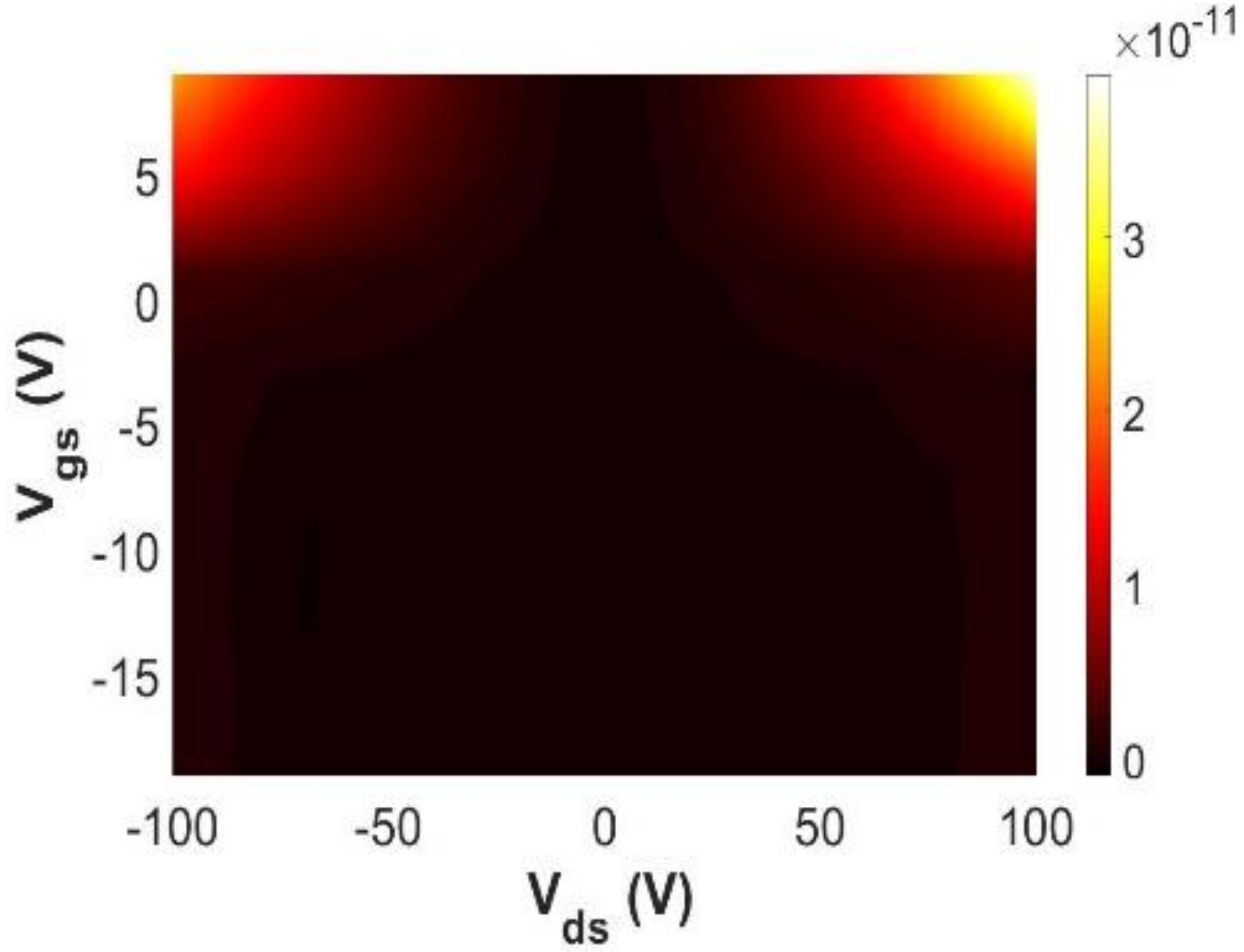

Figure S11. Dissipated Power map *($P_{dark}$)* plotted against different applied $V_{gs}$ and $V_{ds}$.

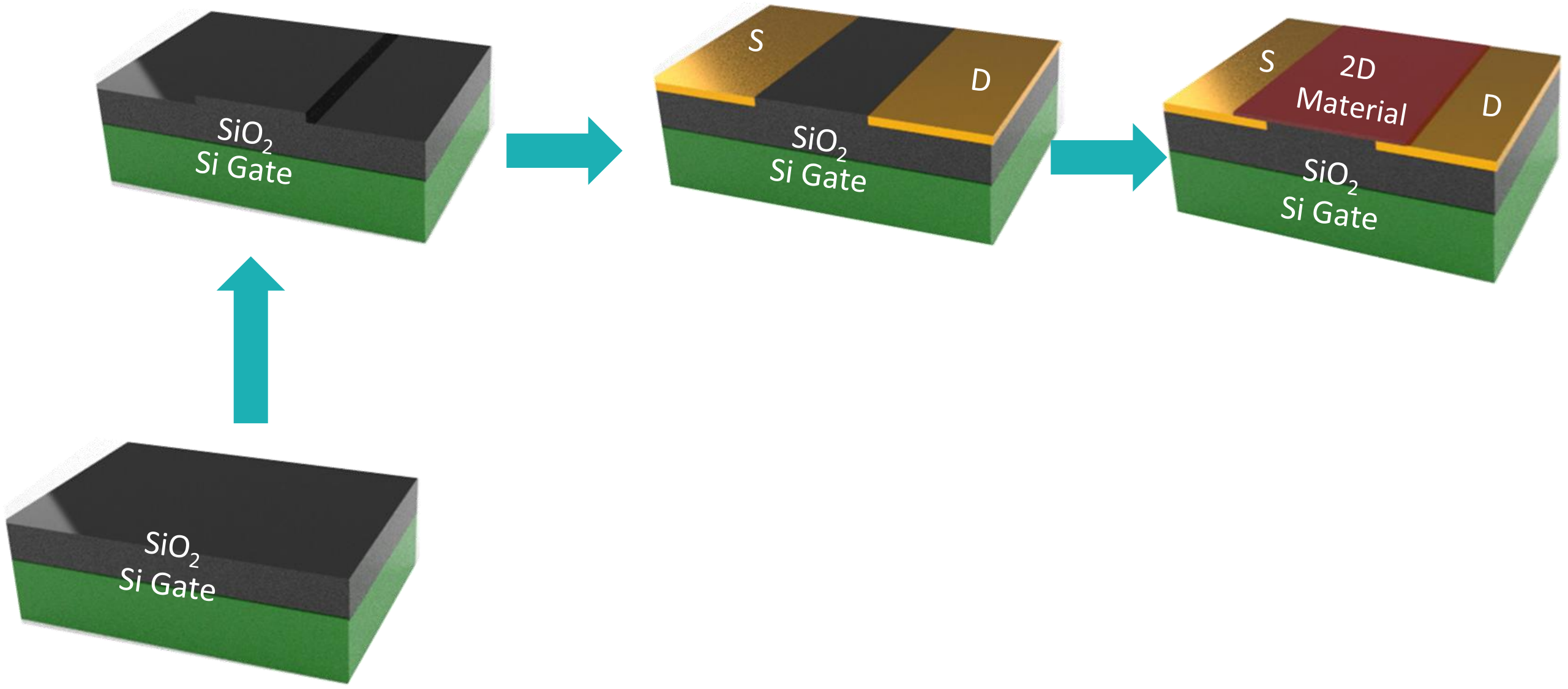

Figure S12. Schematic illustration of the device fabrication process using standard optical lithography followed by metal deposition and finally 2D vdW deposition on metal electrodes.

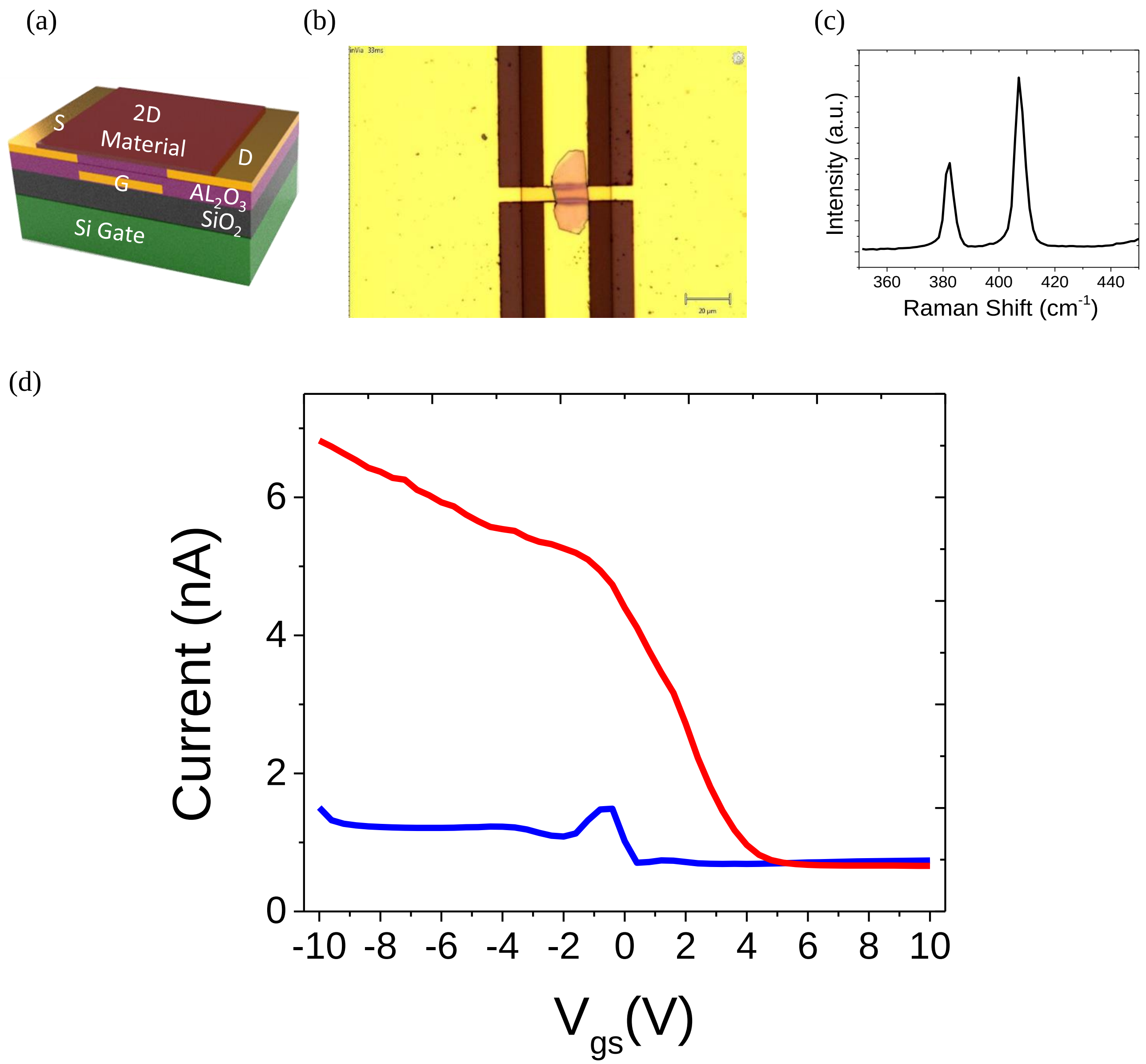

Figure S13. (a) schematic and (b) optical image of $MoS_2$ device fabricated on vdW-FET with single gate (Au) embedded underneath $Al_2O_3$. (c) Raman spectrum showing signatures of $MoS_2$ peaks. (d) $IV_{gs}$ measurements before and after anneal and heal treatment process. The device exhibits pronounced *p*-type behavior instead of *n*-type readily after anneal and heal treatment. This confirms $SiO_2$ contribution to *n*-type doping of $MoS_2$ as reported by other groups. This observed *p*-type is further evidence of weak Fermi Level Pinning at the contacts, making modified vdW deposition lithography an attractive strategy.

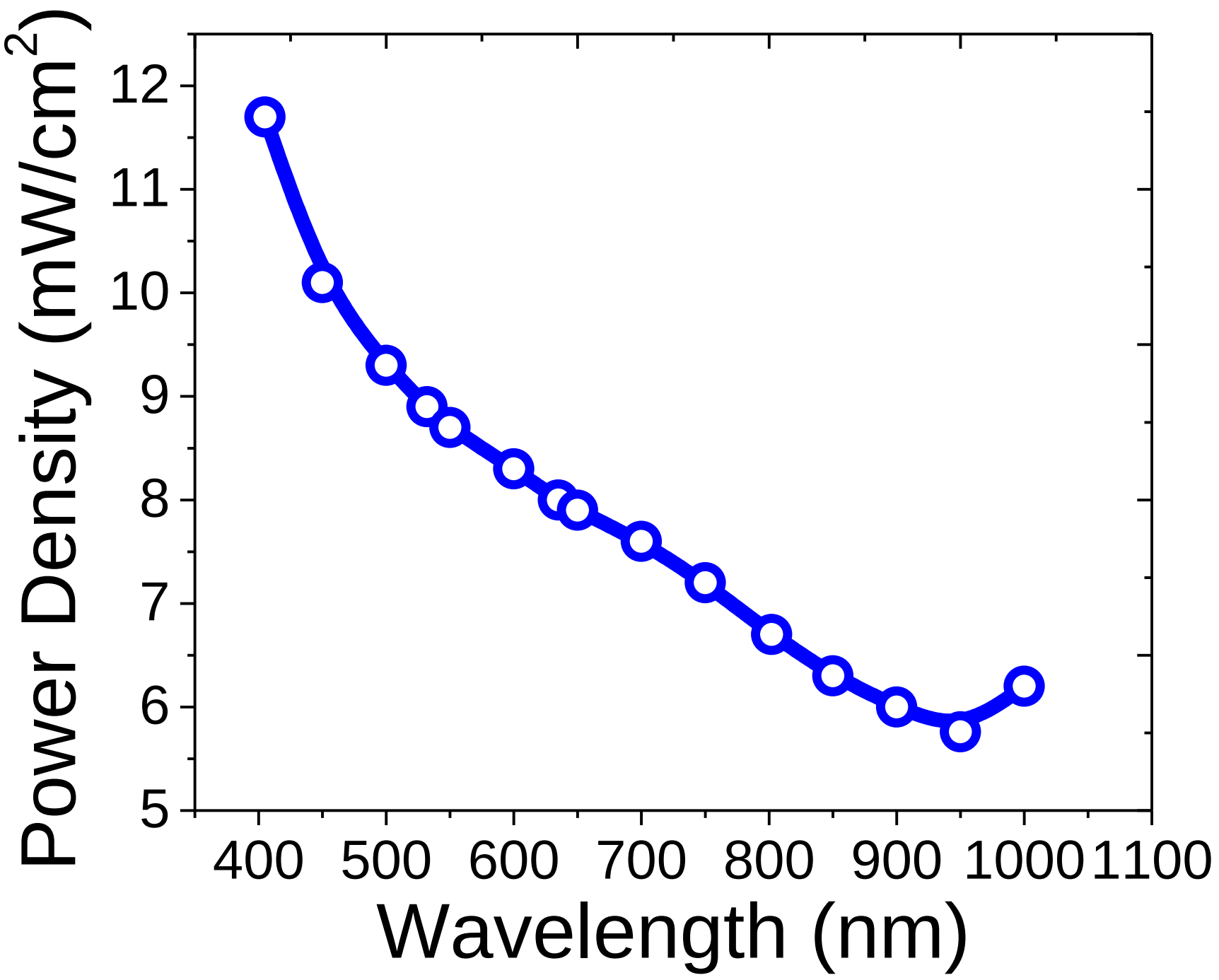

Figure S14. Power density vs. wavelength of white LED used in photodetector measurements. The total power density is given by integrating the area under the curve.